\documentclass[twocolumn]{aastex7}

\usepackage{graphicx}
\usepackage{amsmath}
\usepackage{multirow}
\usepackage{natbib}
\usepackage{amssymb}

\usepackage{caption,subcaption}

\begin{document}

\title{
The Host Galaxies of Fast Radio Bursts Track a Combination of Stellar Mass and Star Formation,
Similar to Type Ia Supernovae
}

\shorttitle{FRB Host Galaxies}
\shortauthors{Horowicz \& Margalit}

\author[orcid=0009-0000-6748-4319 ]{Asaf Horowicz}
\affiliation{Department of Particle Physics and Astrophysics, Weizmann Institute of Science, 234 Herzl St, 7610001 Rehovot, Israel}
\email[show]{asaf.horowicz@weizmann.ac.il}  

\author[orcid=0000-0001-8405-2649]{Ben Margalit}
\affiliation{School of Physics and Astronomy, University of Minnesota, Minneapolis, MN 55455, USA}
\email[show]{margalit@umn.edu}  

\begin{abstract}

We develop a new statistical framework for studying the host galaxies of astrophysical sources
that accounts for both redshift evolution and the multi-variate nature of host-galaxy properties.
These aspects are critical when dealing with sources that span a wide range of redshifts, and/or with unknown redshift-dependent selection effects.
We apply our method to a sample of Fast Radio Burst (FRB) host-galaxies as a means of probing the uncertain progenitor(s) of these events.
Using our method we are able to rule out that FRBs track star-formation rate (SFR), 
as would be expected if FRBs are associated exclusively with young neutron stars born via core-collapse supernovae (SNe). 
Furthermore, we 
rule out a recently proposed metallicity-dependent model whereby FRBs track SFR only above an oxygen abundance of $12 + \log({\rm O}/{\rm H}) \sim 8$.
Motivated by the fact that at least one FRB has been localized to a globular cluster (GC), we also investigate the hypothesis that FRB sources track GC mass and explicitly rule out this scenario. 
Alternatively, we find that a `mixed' model whereby FRBs track a linear combination of both SFR and stellar-mass best explains the data.
The preferred parameters of such a mixed model are similar to those inferred for Type Ia SNe, 
and implies a possible connection between the progenitors of these different transients.


\end{abstract}


\section{Introduction}

Fast radio bursts (FRBs) are $\sim$ms duration luminous radio bursts that were first discovered in archival data \citep{Lorimer+07}. Since then, hundreds more FRB sources detected by dozens of radio telescopes have been published, and this class of mysterious events has led to the emergence of a new active field of research (see e.g., \citealt{Cordes&Chatterjee19,Petroff+22,Bailes22}, for recent reviews).
Despite the large and rapidly growing amount of observational FRB data, the physical mechanism(s) and astrophysical origin(s) of FRB phenomena is still uncertain.

It has long been argued that magnetars---young neutron stars (NSs) with large magnetic fields---are appealing progenitors for powering FRB emission \citep[e.g.,][]{Popov+13,Kulkarni+14,Lyubarsky+14,Katz16,Murase+16,Kumar+17,Metzger+17,Beloborodov17,Lu&Kumar18,Margalit+18,Wadiasingh&Timokhin19,Beloborodov20,Kumar&Bosnjak20}.
These arguments were based primarily on theoretical considerations, as well as circumstantial evidence regarding the local environment and host galaxies of the small number of localized FRBs that were known at the time.
It was only in 2020 that direct and unambiguous evidence supporting the association of FRBs with magnetars was found. This came in the form of a Galactic FRB that was pinpointed to the known galactic magnetar SGR 1935+2154 \citep{Andersen+20,Bochenek+20}. This watershed event confirms that---at the very least---a subset of extragalactic FRBs can be produced by magnetars, although the homogeneity and formation channels of such magnetar progenitors is still uncertain \citep[e.g.,][]{Margalit+20,Lu+20}.

One challenge to the naive magnetar scenario is the environment of some FRB sources. With the growing sample of localized FRBs whose host galaxies have been identified, it has been found that some FRBs reside in quiescent galaxies with very little recent star-formation \citep[][]{Bannister+19,Ravi+19,Connor+23,Law+24,Shah+25,Eftekhari+25}. This is in tension with the conventional expectation for magnetars---that these sources should be formed through the core-collapse of massive stars. In this scenario, FRB sources should closely-track star-formation and their occurrence rate in quiescent galaxies should be fairly low.\footnote{Although, as shown by \cite{Loudas+25}, a small fraction of FRBs would be expected to reside in quiescent galaxies even in scenario where FRBs track star-formation \cite[see also][for similar results in the context of core-collapse SNe]{Nugent+25}.}
More direct evidence along this line has recently surfaced with the localization of a nearby FRB to a globular cluster (GC; \citealt{Kirsten+22}). Redshift zero GCs are characterized by old stellar populations with no ongoing star-formation, directly  challenging standard magnetar formation channels.
In light of these observations alternative formation channels for magnetar FRB sources have been proposed, including the possibility that magnetars could be formed in the accretion-induced-collapse (AIC) of a white dwarf (WD), or in the aftermath of a subset of binary NS mergers \citep{Margalit+19,Kremer+21,Lu+22,Eftekhari+25}.
Questions pertaining to the sources of FRBs, and whether some or all of them are related to magnetars (and if so, what are the formation channels and characteristics of such magnetars) are therefore central to our understanding of FRB phenomena.

A particularly powerful tool in addressing these questions is studying the population of host-galaxies and local environments of FRB sources \citep{Tendulkar+17,Bannister+19,Bochenek+20,Heintz+20,Mannings+21,Bhandari+22,Gordon+23,Bhardwaj+24,Woodland+24,Sharma+24,Loudas+25,Gordon+25,Shah+25,Eftekhari+25}.
Although host-galaxy studies in general offer only circumstantial evidence about the nature of sources, the ability to statistically study large samples via this approach makes this method appealing. Indeed, host-galaxy studies have been central to building an understanding of other transient phenomena. They have helped solidify the association of (at least typical) long gamma-ray bursts (GRBs) with low-metallicity star-forming environments, consistent with expectations from the collapsar model \citep{Fruchter+06}; they have shown that superluminous supernovae (SLSNe) occur in similar environments to long GRBs, potentially consistent with rapidly-rotating magnetar sources \citep{Lunnan+14}; and they have shown that the formation channels of short GRB and Type Ia supernovae (SNe) progenitors differ significantly from other transients, tracking instead a combination of both star-formation and stellar mass \citep[e.g.,][]{Scannapieco+05,Fong+13,Nugent+22,Nugent+25}.

The study of FRB host galaxies has been a particularly-active sub-field in recent years. Such studies have, until recently, been limited by the relatively small number of localized FRBs. However, technical advances and intense instrumental and observational efforts are starting to change this landscape: currently there are of order $\sim$100 published FRB host-galaxies, up from only a handful a few years back. This sample size is expected to grow exponentially in the near future, with more precise localizations coming from 
the Deep Synoptic Array (DSA-110),
the Canadian Hydrogen Intensity Mapping Experiment's (CHIME) outriggers program,
and the Australian Square Kilometre Array Pathfinder's (ASKAP) Commensal Realtime ASKAP Fast
Transient COherent (CRACO) upgrade
\citep[e.g.,][]{Ravi+23,Lanman+24,Wang+25}.
It is therefore timely to re-examine the methodology and framework through which these data may be interpreted.

In this paper, we present a new framework for studying the host galaxies of astrophysical sources, and apply this framework to the existing sample of FRB host-galaxies.
We begin by summarizing past work in this space, and discussing general considerations and limitations on host-galaxy studies involving astrophysical sources that span a wide range of redshifts (\S\ref{sec:past_work}). In \S\ref{sec:methods} we introduce the formalism of our new framework. We then apply our model to the current sample of FRB host-galaxies in \S\ref{sec:results}, and discuss our findings and their implications. Finally, we conclude and summarize our main results in \S\ref{sec:discussion}.

\section{Past Work}
\label{sec:past_work}

Statistical studies of the host-galaxies of transient events take many forms. Most commonly, `bulk' properties of the host galaxies (i.e., properties that describe the entire galaxy rather than the more local environment of the transient within the galaxy) are considered. Chief among these are the galaxy stellar mass $M$ and star-formation rate (SFR), though other properties such as metallicity are also commonly considered. These properties are then compared to some reference sample of galaxies. This might be the host galaxies of some other class of transient, or the distribution of field galaxies from a galaxy survey.
In either case, the statistical comparison between the two samples typically aims to answer the question: {\it are the two galaxy samples consistent with being drawn from the same underlying distribution?}

The prevailing means of addressing this question are the use of a Kolmogorov-Smirnov (KS) or Anderson-Darling (AD) test. These statistical methods examine the 1D cumulative distribution of two samples with respect to some variable (e.g., the mass or SFR of galaxies), and test the hypothesis that the two samples are drawn from the same underlying distribution.
These methods are well established and have played an important role in improving our understanding of transients \citep[e.g.,][]{Lunnan+14,Heintz+20,Taggart&Perley21,Schulze+21,Bhandari+22,Nugent+22,Law+24,Sharma+24}.
However, it is important to note that from a mathematical perspective, 
KS and AD tests are only strictly valid for univariate distributions.

In contrast, host-galaxy properties are intrinsically multivariate and characterized by a combination of several parameters, such as $M$, SFR, and redshift (at the very least).
To use KS or AD tests, host-galaxy samples are therefore typically ``marginalized'' over a number of variables so that the distribution of some single variable (say stellar mass) may be examined.
In addition to the convenience of being able to use 1D KS tests, such marginalization also has merit in cases where some variables are more robust than others (indeed, recovering SFRs is significantly more challenging than correctly recovering stellar masses).
This approach, though useful, is not statistically sound. 

For example, it is possible for two intrinsically different galaxy distributions in the $(M,{\rm SFR})$ plane to have similar 1D distributions in $M$ and/or SFR. Univariate KS or AD tests might therefore suggest that the two samples are drawn from the same distribution, when in fact they are not. 
One way to conceptualize this ambiguity is to consider that the choice of axes along which to marginalize is arbitrary in a multivariate phase-space. In particular, there is no apriori reason why KS tests should be run on the masses and SFRs of galaxies rather than on some alternative linear combination of the two. The fact that this arbitrary choice of axes will affect the KS score and, perhaps, the resulting conclusion is therefore troubling.
Unfortunately, there is no distribution-free generalization of the KS or AD tests to more than one dimension (although see e.g., \citealt{Peacock83} and \citealt{Fasano&Franceschini87} for KS generalizations that are only weakly-dependent on the assumed distribution).\footnote{`Distribution-free' means that there is no statistical method to test whether two multivariate samples are drawn from the same underlying distribution, without making any assumptions about the functional form of the underlying distribution.}

A further complication arises in considering the role of redshift in host-galaxy studies. This is particularly important when dealing with transients that span a wide range in $z$, such as FRBs.
In principle redshift can be treated as just another variable in the galaxy distribution function, on equal footing with $M$ and SFR. However in practice, there may be reasons to avoid this approach. Selection effects are one reason, as transient host-galaxy samples are rarely complete. Another is that the comparison (reference) galaxy sample may not span a suitable redshift range, limiting this approach.
Regardless, accounting for redshift evolution is crucial for transients that span an appreciable range in $z$. 

This was first pointed out in the context of FRBs by \cite{Bochenek+21}.
In that work, the authors devised a method of `transposing' host galaxies to redshift $z=0$. They then compared the transposed host galaxy distribution with a complete sample of core-collapse SN (CCSN) hosts at $z=0$ and found that the two samples are consistent with one another, but only after accounting for this redshift scaling (transposition).
This illustrated the importance of accounting for redshift when making such comparisons, which can affect the overall conclusions markedly.
However the particular methodology presented by \cite{Bochenek+21}---though novel and important---has one noticeable shortcoming. 
In that work, the stellar mass and SFR of each FRB host-galaxy were rescaled to $z=0$ based on the star-forming main sequence of galaxies. Specifically, each galaxy's stellar-mass and SFR were transposed in a way that preserves the
galaxy's percentiles in both $M$ and SFR {\it relative to the distribution of star-forming galaxies} at the redshift of the transient and at $z=0$.
Implicit in this choice is the assumption that FRB hosts evolve according to the redshift evolution of star-forming galaxies. However, there is no guarantee apriori that this be the case. Indeed, if one is interested in testing the hypothesis that FRBs track anything other than SFR, this choice may bias the results.
An alternative method that avoids `transposition' is presented by \cite{Sharma+24} (see their Equation~11). The authors choose a specific property of host galaxies and define a test statistics that evaluates how extreme an FRB host galaxy is compared to the background population at its redshift. While sidestepping the need for `transposing' galaxies to redshift $z=0$, this method operates on only one property at a time, losing potential information encoded in the multivariate nature of host galaxies.

Another method that is commonly employed to account for redshift evolution is binning. In this approach, FRB host galaxies are separated into a number of subsamples based on their redshift. The FRB hosts in each redshift bin are then compared to a reference galaxy distribution spanning the same, limited, range in redshift \citep[e.g.,][]{Heintz+20,Bhandari+22,Gordon+23,Sharma+24,Loudas+25}.
This method does not make any assumptions about the underlying redshift evolution and is therefore better suited for model-independent tests. However, it is limited by the particular choice, number, and size of redshift bins used. In particular, with a small sample size of FRB hosts, fairly wide redshift bins must be used in order to have a statistically-meaningful number of galaxies in any bin. This will improve in the near future as the number of well-localized FRBs increases.
However even so, this approach inevitably dilutes the statistical power of the full host-galaxy sample by comparing smaller subsamples at each redshift.
Furthermore, if the redshift bins are sufficiently wide then the effects of redshift evolution are again important because the population and the observational selection effects may change within the span of a single bin.

In the following, we introduce an alternative approach that avoids some of the pitfalls discussed above.
In particular, the methodology that we describe in \S\ref{sec:methods} accounts for redshift evolution in a self-consistent model-independent manner, that allows one to use the full statistical power of a large sample of host-galaxies. Furthermore, our methodology relies on a sampling of the multivariate galaxy property phase-space (in this case, $M$ and SFR) in a statistically robust manner that does not require marginalization into univariate distributions.
A trade-off of this approach is that our method provides only a conservative test for ruling out model hypotheses. That is, it is theoretically possible to construct two distributions that are very different from one another, but that would produce a high p-value for being drawn from the same underlying distribution.
Conversely however, if two distributions are found to have a low p-value using our method, then the conclusion that they are not drawn from the same underlying distribution is robust.
In this sense, our method should be used as a test for ruling out the hypothesis that two galaxy distributions are consistent with one another.

\section{A New Framework}
\label{sec:methods}

Given a sample of $i=1,..,N$ galaxies 
with measured stellar masses $M_i$, current star-formation rates ${\rm SFR}_i$, and redshifts $z_i$,
we wish to test the hypothesis that these $N$ galaxies are consistent with being drawn from some model distribution. 
In the following section we describe the formalism with which we test this hypothesis.

We use a redshift-dependent galaxy distribution function derived by \citet{Leja+22} from spectral energy distribution (SED) modeling of galaxies using the non-parametric {\tt Prospector} code \citep{Leja+17,Johnson+21}.
This continuous distribution function provides the probability density $\rho(M, {\rm SFR}, z)$ of randomly drawing a galaxy that has stellar mass $M$, star-formation rate SFR, and redshift $z$. It is trained on the 3D-HST and COSMOS-2015 galaxy surveys \citep{Skelton+14, Laigle+16} and can therefore be viewed as an empirical distribution function.

One important caveat in adopting this model is the necessity that properties of the comparison galaxy sample (in this case localized FRB host galaxies) be derived in a consistent manner---using the same galaxy SED-fitting code and assumptions. 
\cite{Leja+19} show that using different codes and/or assumptions can systematically affect inferred galaxy properties. In particular, using {\tt Prospector} and a non-parametric star-formation history, \cite{Leja+19} found that the stellar masses and SFRs of galaxies are on average 0.1--0.3 dex larger and 0.1--1 dex lower (respectively) than previous studies. The largest contributing factor to these differences was attributed to the use of non-parametric star-formation histories.
We therefore use the FRB host-galaxy sample presented in \cite{Sharma+24}, which builds on previous works by \cite{Gordon+23} and \cite{Bhardwaj+24}, and includes $N=51$ FRB host galaxies whose properties are derived using the {\tt Prospsector} code, consistently with \cite{Leja+22}.
See \S\ref{sec:sample} for more details.

\subsection{Extending the Distribution Function}
\label{sec:continuity}

Being trained on galaxy catalogs, the \cite{Leja+22} distribution function is limited by these catalog's range and completeness. In particular the distribution is valid between redshifts $0.2 < z < 3$, and is only complete above a redshift-dependent mass-completeness limit $M_{\rm c}(z)$.
The latter is due to lower mass galaxies being fainter and harder to detect in surveys. Of course, even though galaxy catalogs may not be complete to such galaxies, galaxies of mass $M < M_{\rm c}(z)$ are in reality abundant. 
In the FRB host-galaxy sample that we adopt there are two sources (out of 51) that fall below the threshold mass described above.
To take these two galaxies into account, and to allow our formalism to be applied more broadly without limitation on galaxy masses, we extend the \cite{Leja+22} distribution below $M_{\rm c}$ using the following two considerations.

First, we use the star-forming sequence. When the distribution of galaxies is plotted as a function of galaxy stellar mass and SFR, there is a well-known concentration of galaxies along a curve called the star-forming sequence (or sometimes, the star-forming main sequence). 
We extend the probability density function below some threshold mass $M_{\rm th}$ such that
each lower-mass point in the distribution maps to a point at $M=M_{\rm th}$ that is an equal distance (along the SFR-axis) from the star-forming sequence.
Second, we account for the increasing number density of low mass galaxies by adding a correction term that is proportional to the stellar mass function, $\Phi(M,z)$.
Combining these considerations, we extend the probability density function $\rho$ below the mass-completeness threshold as follows,
\begin{align}
    \label{eq:rho_extension}
    \rho \left( M<M_{\rm th} ,\, {\rm SFR} \,\vert\, z \right) 
    = \rho &\left[ M_{\rm th} ,\,  \left( {M}/{M_{\rm th}} \right)^{-b(z)} {\rm SFR} \,\vert\, z \right]
    \nonumber \\ 
    &\times {\Phi(M,z)}/{\Phi(M_{\rm th},z)}
    .
\end{align}
Here $b(z)$ is the (redshift-dependent) low-mass slope of the star-forming sequence as given by \cite{Leja+22},\footnote{
The parameter $b$ is defined in Equation~9 of that paper. Its dependence on redshift is parameterized via their Equation~10, with best-fit coefficients listed in their Table~1.
Here we adopt parameters as fit to the `ridge' of the star-forming sequence.}
and $\Phi(M,z)$ is a double Schechter-function fit to the stellar mass function taken from \cite{Leja+20}.
Finally, we take $M_{\rm th}= 2 M_{\rm c}(z)$ where $M_{\rm c}(z)$ is the mass-completeness cutoff described in \cite{Leja+22} (e.g., their Table 1). We take a threshold mass that is a factor of two larger than $M_{\rm c}$ to avoid possible edge-effects that may affect the distribution near the mass-completeness cutoff.

The extension described in Equation~(\ref{eq:rho_extension}) effectively assumes that low mass galaxies with $M<M_{\rm th}$ follow the star-forming sequence (or more accurately, that they can be transposed with relation to the star-forming sequence, even if they fall far from its locus). In general, this is a strong assumption. Many galaxies are quiescent or transitioning and do not follow the star-forming sequence. Given that we are interested in testing the hypothesis that FRB hosts track SFR and/or stellar mass, it is crucial to account for these quiescent galaxies. Indeed, our present work is motivated by shortcomings of previous studies in this respect. Ideally we would not need to extend the distribution function in this way. However we must define some extension if we wish to include the two FRB hosts that fall below the mass-completeness threshold. 
We argue that Equation~(\ref{eq:rho_extension}) is nevertheless a reasonable choice for the following reasons.
First, low-mass galaxies---those that are affected by our extension---have a strong preference towards being star-forming, while quiescent galaxies (which do not track the star-forming sequence) mostly have $M>M_{\rm th}$. In other words, using the star-forming sequence is a reasonable approximation for low mass galaxies, exactly the regime where Equation~(\ref{eq:rho_extension}) is relevant.
Second, there are only two sources in our FRB sample that are affected by this extension, and both lie only slightly below the cutoff mass. That is, the extrapolation introduced by our extension is very small, and only applies to a small fraction of our sample (see Table~\ref{tab:mass_cutoff}).

One final extension we apply to the \cite{Leja+22} distribution function is a treatment that incorporates low-redshift $z<0.2$ galaxies, where the \cite{Leja+22} distribution is truncated. Ideally, a model that is trained on low-redshift galaxy catalogs would be incorporated within this framework and allow a self-consistent evolution with redshift. However this is well beyond the scope of our present work. In order to include low-redshift FRB host galaxies in our current analysis we therefore extend the \cite{Leja+22} distribution in redshift-space using the following approximation: we use the extrapolated $z<0.2$ redshift dependence implied by fitting functions to the star-forming sequence and the stellar-mass function obtained by \cite{Leja+19,Leja+22}. These are extrapolations because the fits were obtained on data at $z> 0.2$. In practice, this implies the following extension,
\begin{align}
\label{eq:z_extension}
    \rho \left( M,\, {\rm SFR} \,\vert\, z<z_{\rm th} \right) 
    = \rho &\left[ M,\,  
    \frac{\langle{\rm SFR}\rangle(M,z_{\rm th})}{\langle{\rm SFR}\rangle(M,z)} 
    {\rm SFR} \,\vert\, z 
    \right] 
    \nonumber \\
    &\times {\Phi(M,z)}/{\Phi(M,z_{\rm th})}
    .
\end{align}
Here $\langle{\rm SFR}\rangle(M,z)$ is the `ridge' of the star-forming main sequence, as modeled by \cite{Leja+22} (see their Equation~9).
In Equation~(\ref{eq:z_extension}) we take $z_{\rm th}=0.25$ as a cutoff redshift instead of $z=0.2$ in order to avoid possible edge-effects of the distribution near the minimal redshift. 

Our redshift extension may be overly simplified, however it is born out of necessity given that we wish to consider FRB host galaxies that would otherwise be excluded from our analysis given the redshift cutoff. 
Overall, our approach to extending the \cite{Leja+22} distribution function to low redshifts and low galaxy masses is very similar to the approach recently taken by \cite{Sharma+24} and \cite{Loudas+25}.

\subsection{Weights and Models}
\label{sec:likelihood}

\begin{figure*}
    \centering
    \begin{subfigure}[b]{0.3\textwidth}
        \centering
        \includegraphics[width=\textwidth]{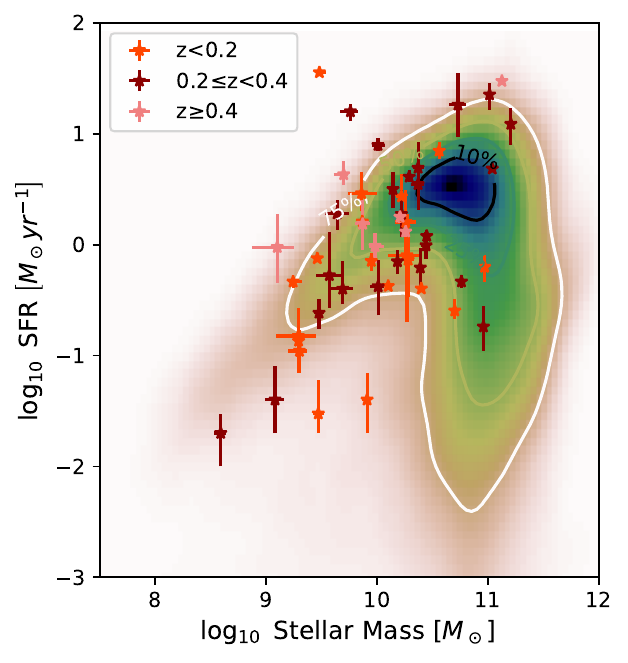}
        \caption{Mass-weighted}
    \end{subfigure}
    \hfill
    \begin{subfigure}[b]{0.3\textwidth}
        \centering
        \includegraphics[width=\textwidth]{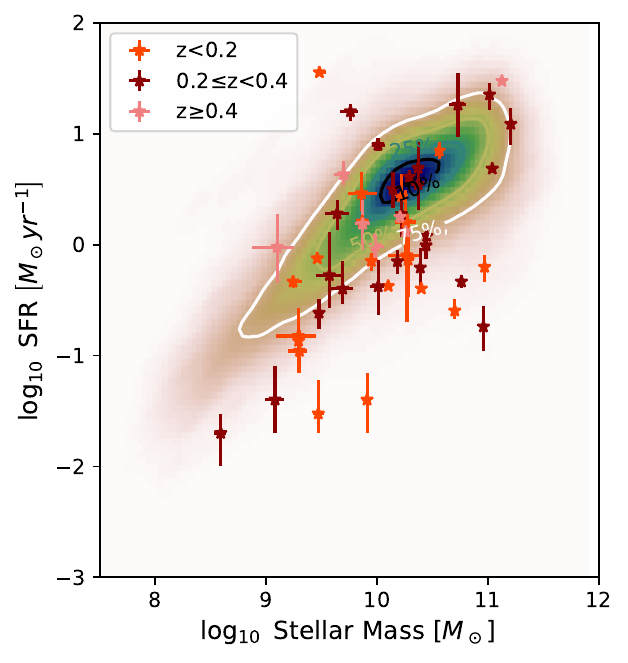}
        \caption{SFR-weighted}
    \end{subfigure}
    \hfill
    \begin{subfigure}[b]{0.3\textwidth}
        \centering
        \includegraphics[width=\textwidth]{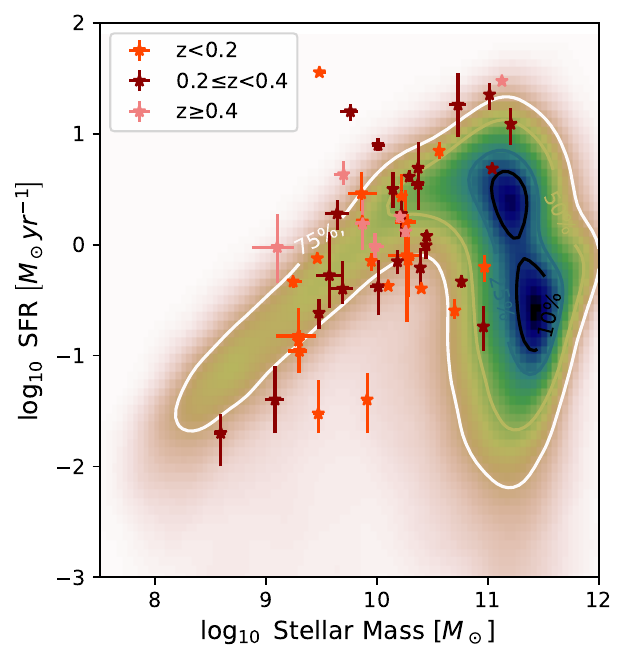}
        \caption{GC-weighted}
    \end{subfigure}
    \caption{
    Weighted galaxy distributions in the stellar-mass--SFR plane, illustrated here at a fixed redshift of $z=0.3$. 
    The unweighted distribution of field galaxies $\rho \left( M,{\rm SFR} | z \right)$ is based on a continuous model from \cite{Leja+22}.
    Panels (a), (b), and (c), show the mass-weighted, SFR-weighted, and GC-weighted distributions, respectively, as defined via Equations~(\ref{eq:weights_mass},\ref{eq:weights_SFR},\ref{eq:weights_GC}).
    In each panel, black, yellow, and white contours show regions that enclose 10\%, 50\%, and 75\% of the probability.
    A $z=0.3$ sample of FRB host-galaxies would follow one of these distributions if FRBs tracked stellar mass, SFR, or the mass of stars in GCs.    
    Colored points show the FRB host-galaxy sample considered in this paper (\S\ref{sec:sample}; \citealt{Sharma+24}) for illustrative comparison purposes. As these galaxies span a range of redshifts, they cannot be directly compared to the background distributions which are at fixed $z=0.3$. Our analysis takes this into account by comparing each FRB host to field galaxies at its redshift.
    }
    \label{fig:weighted_dists}
\end{figure*}

We define the likelihood of a sample of $i=1,..,N$ galaxies as
\begin{equation}
\label{eq:likelihood}
    \mathcal{L} = \sum_{i=1}^{N} \ln \left[ W_i \rho \left(M_i, {\rm SFR}_i \vert z_i \right) \right]
    ,
\end{equation}
where $\rho$ is the probability density function of galaxies 
that was presented previously in this section, 
and $W_i$ are model-dependent weights that we discuss shortly.
Beforehand, we note that our definition above intentionally uses $\rho \left(M_i, {\rm SFR}_i \vert z_i \right)$ rather than $\rho \left(M_i, {\rm SFR}_i, z_i \right)$. That is, we evaluate the probability of each galaxy being drawn from the underlying distribution given its measured redshift. Stated another way, we avoid weighting galaxies based on their redshift, and only treat the stellar mass and SFR as independent variables. This is because the redshift of our FRB host-galaxy sample is almost certainly biased. In principle this restriction can be lifted if dealing with an unbiased volume-limited galaxy sample and redshift evolution is an important component of models that one wants to test. However we avoid this approach in our present work given limitations of current FRB host-galaxy samples.\footnote{
Note that such biases cannot be easily corrected, and extend beyond host-galaxy magnitude selection effects (which we do account for, following \citealt{Sharma+24}). This is because the FRB host-galaxy sample depends also on the radio selection function of FRBs. 
Given uncertainties about the FRB luminosity function, repetition rate, etc., such effects cannot be easily quantified at present.
}

The weights $W_i$ encode information regarding the underlying model that we wish to test. For example, the trivial case where $W_i = const.$ describes a model whereby FRBs are equally-likely to occur in any kind of galaxy and should therefore follow the distribution of field galaxies in the Universe.
A more physically-motivated choice of weights might be parameterized as
\begin{equation}
\label{eq:weights_mass}
    \text{`mass-weighted':}~~~~~
    W_i
    = M_i 
    ,
\end{equation}
\begin{equation}
\label{eq:weights_SFR}
    \text{`SFR-weighted':}~~~~~
    W_i
    = {\rm SFR}_i 
    ,
\end{equation}
or
\begin{equation}
\label{eq:weights_AB}
    \text{`mixed':}~~~~~
    W_i 
    = A \left(\frac{M_i}{M_\odot}\right) + B \left(\frac{{\rm SFR}_i}{M_\odot / {\rm yr}}\right)
    .
\end{equation}
Equation~(\ref{eq:weights_AB}) allows for a linear combination of dependence on galaxy stellar mass and SFR. 

For $A=0$ this reduces to a SFR-weighted distribution (Equation~\ref{eq:weights_SFR}). This might be expected if FRBs, like CCSNe, were produced only by young stellar populations.
Conversely, for $B=0$ Equation~(\ref{eq:weights_AB}) reduces to a mass-weighted distribution (Equation~\ref{eq:weights_mass}) that can be interpreted as a preference for older stellar populations. Weighting by mass alone ($B=0$) can also be viewed as a proxy for weighting by the total number of stars (assuming some initial-mass function). Therefore, another possible interpretation of this scenario would be that FRBs can be produced by any kind of star and are therefore more likely to occur in galaxies that have a larger number of stars.
Note that the absolute values of the constants $A$ and $B$ are only important if one wants to determine an overall volumetric rate. We are not interested in this question 
in our present work, and therefore only the ratio of $A/B$ will be important.

An alternative choice of weights that we consider in this work is motivated by the association of one FRB source to a globular cluster (GC; \citealt{Kirsten+22}). If all FRBs occurred in GCs then one estimate of their host-galaxy preference might be encoded through the total GC mass per galaxy, $M_{\rm GC}$. The associated weighting-function is then
\begin{equation}
\label{eq:weights_GC}
    \text{`GC-weighted':}~~
    W_i = M_{\rm GC}(M_i)
    = \eta_{\rm GC} M_{\rm halo}(M_i)
    .
\end{equation}
We calculate the total mass in GCs for a given galaxy assuming that it is a fixed fraction $\eta_{\rm GC}$ of the galaxy's halo mass $M_{\rm halo}$, 
and take $\eta_{\rm GC} = 10^{-4.54}$.
This proportionality has been shown to hold reasonably well (\citealt{Harris+13,Hudson+14}; except perhaps for very low mass galaxies, e.g., \citealt{Eadie+22}), and we take the constant of proportionality from \cite{Harris+17}. To estimate the halo mass of a galaxy $M_{\rm halo}(M)$ we use the stellar-mass--halo-mass function from \cite{Behroozi+19}.
Because of the non-linearity of the stellar-mass--halo-mass relation, Equation~(\ref{eq:weights_GC}) depends non-linearly on the galaxy stellar mass $M$.
We refer to this scenario where FRBs track the mass of GCs as the `GC-weighted' model.

This GC-weighted model is probably an over-simplification in that we treat the total mass of stars in GCs as a proxy for the FRB occurrence rate. To the extent that GCs are a common formation channel for FRB sources, it is likely that dynamical interactions are crucial for their formation \citep[e.g.,][]{Kremer+21,Lu+22}. This would imply that the interaction rate of a GC is more direct measure of the probability of an FRB source occuring in a GC, rather than the total mass \citep[e.g.,][]{Pooley+03}.
Given that we are unaware of any published relationships between the GC interaction rate and host galaxy properties (which are the measurable quantities in our sample), we were forced to consider the simplified GC mass model. However we encourage future investigations along these lines.

\subsection{FRB Host-galaxy Sample}
\label{sec:sample}

We choose to work with the FRB host-galaxy samples presented in 
\cite{Gordon+23}, \cite{Bhardwaj+24}, and
\cite{Sharma+24}, the largest published data set (at the time writing our work) with galaxy properties inferred using {\tt Prospector} SED-modeling code. As previously discussed, because of systematic differences between galaxy SED-modeling codes, it critical that both the FRB host-galaxy sample and the comparison galaxy sample (in our case the distribution derived by \citealt{Leja+22}) be modeled uniformly using the same code and model assumptions.
The FRB host-galaxy sample used here includes 23 hosts localized by ASKAP \citep{Gordon+23}, 4 hosts found by CHIME \citep{Bhardwaj+24}, and 26 hosts from DSA-110 with secure associations \citep{Sharma+24}. All of these galaxy's properties were derived using the {\tt Prospector} code and follow selection criterion presented in \cite{Sharma+24}. 
Following \cite{Sharma+24} we also take into account an optical selection bias for galaxies, and generate mock data (see \ref{sec:model_testing}) that only includes galaxies whose $r$-band magnitude is $\le 23.5$. 
We follow the methodology presented in \cite{Sharma+24}
in estimating the $r$-band magnitude of mock galaxies
using a prescribed mass-to-light ratio (\citealt{Sharma+24}; see ``Background Galaxy Population'' section in that work).

We note that, in the final stages of writing our manuscript, a paper by \cite{Loudas+25} came out which performed a similar analysis to ours in several respects (although with some important differences).
One thing worth noting at this point is that \cite{Loudas+25} develop a more detailed treatment for estimating mock galaxy $r$-band magnitudes. We leave it to subsequent work to implement this more detailed prescription into our framework. \cite{Loudas+25} find that this improved treatment of the optical selection bias is enough to impact conclusions regarding consistency of the FRB host galaxy sample with star-formation. Subtle differences between these approaches may therefore have an appreciable impact on the derived results, and this is an important caveat to keep in mind.

In total, our primary sample consists of $N=51$ FRB host galaxies.
When presenting our results (\S\ref{sec:results}), we also consider a more conservative analysis of a subset of $N=23$ galaxies. This conservative set excludes galaxies that fall below the mass-completeness threshold or the minimal redshift cutoff of the \cite{Leja+22} comparison distribution, and therefore avoids the need for any extension of this distribution, as described in \S\ref{sec:continuity}. 
In the Appendix we list the two FRBs that fall below the mass-completeness threshold (Table~\ref{tab:mass_cutoff}), 
and the 27 FRB hosts in our sample with redshifts $z<0.25$ that are affected by the redshift extension (Table~\ref{tab:low_z}).
Note that one of these sources (FRB~20210117A) has both low redshift and low mass and is therefore listed in both tables. There are therefore 28 sources 
(out of a total of 51) 
that are affected by the extensions discussed in \S\ref{sec:continuity}.

\subsection{Model Testing}
\label{sec:model_testing}

\begin{figure*}
    \centering
    \begin{subfigure}[b]{0.3\textwidth}
        \centering
        \includegraphics[width=\textwidth]{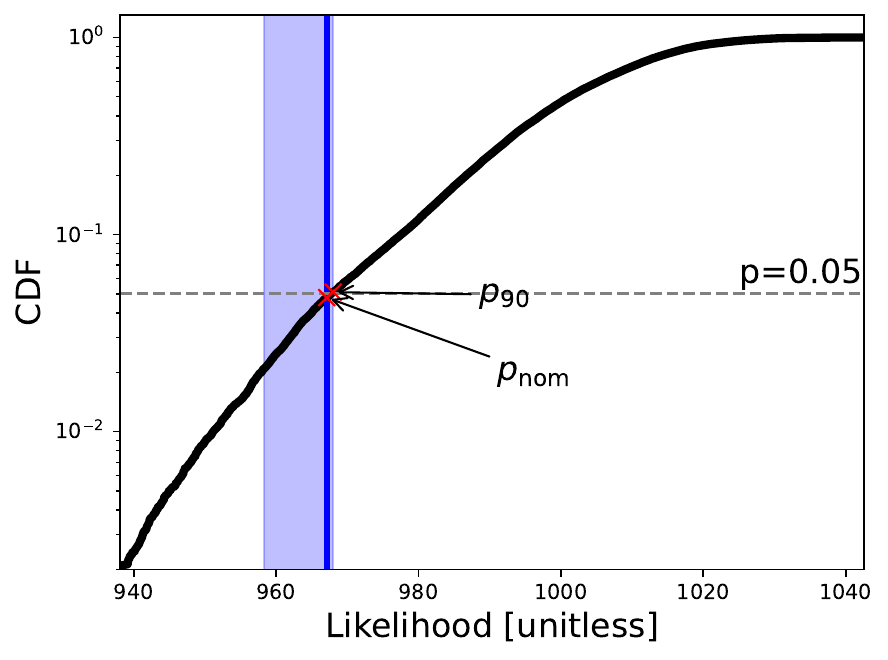}
        \caption{Mass-weighted}
    \end{subfigure}
    \hfill
    \begin{subfigure}[b]{0.3\textwidth}
        \centering
        \includegraphics[width=\textwidth]{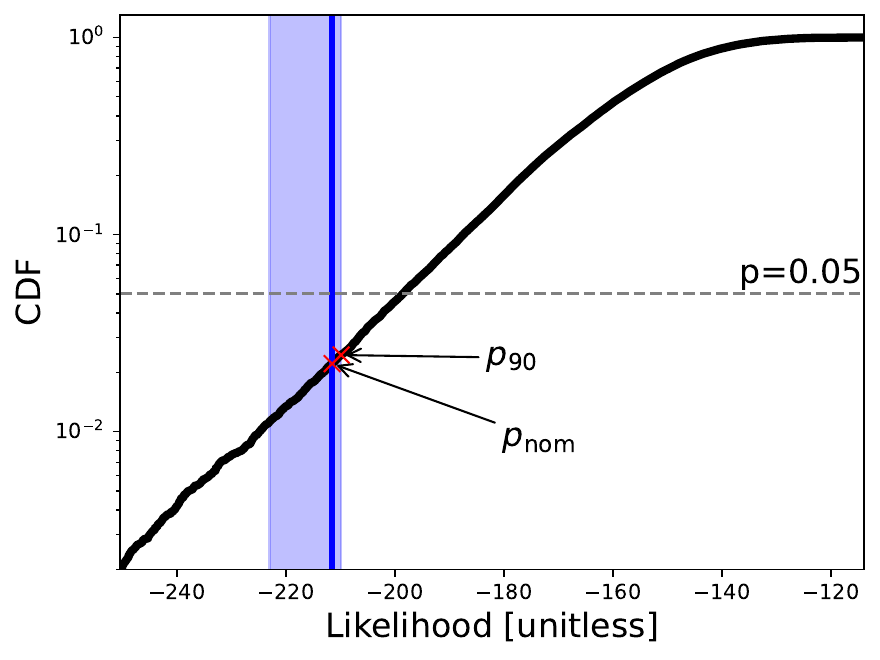}
        \caption{SFR-weighted}
    \end{subfigure}
    \hfill
    \begin{subfigure}[b]{0.3\textwidth}
        \centering
        \includegraphics[width=\textwidth]{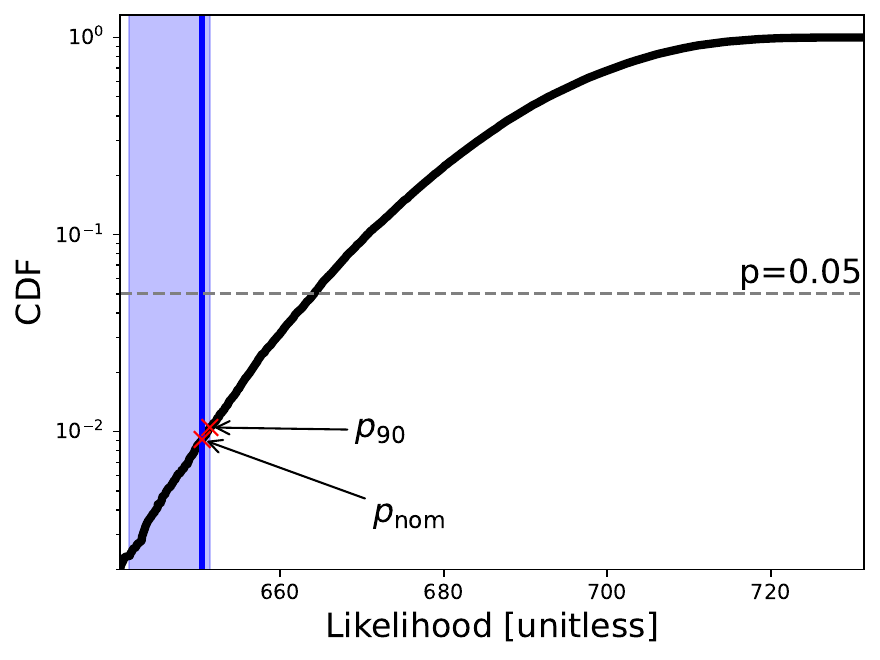}
        \caption{GC-weighted}
    \end{subfigure}    
    \caption{
    Methodology for calculating the p-value of models. Random samples are drawn from a model distribution, effectively generating many mock data sets. The likelihood of every mock dataset is then calculated via Equation~(\ref{eq:likelihood}). Black curves in each panel show the cumulative distribution of likelihoods for the mock data sets. This is compared to the likelihood obtained for the observed FRB host-galaxy sample, shown in blue. The vertical dark-blue curve shows the likelihood obtained using the nominal (median) inferred masses and SFRs for the 51 FRB host galaxies in our full sample. The light-blue shaded region shows the range of values that can be obtained accounting for the uncertainties on the inferred masses and SFRs.
    The nominal p-value of a given model, $p_{\rm nom}$, is defined 
    via Equation~(\ref{eq:pval}) and is found
    by the value of the CDF at the intersection of the dark-blue curve with the black curve. 
    Similarly, a more conservative $p_{90}$ p-value is defined by the intersection of the black curve with the upper bound of the shaded light-blue region.
    A model is considered to be ruled out if the p-value is less than $p=0.05$. 
    }
    \label{fig:pvals}
\end{figure*}

Having defined the likelihood $\mathcal{L}$ of a sample of $N$ FRB host galaxies in Equation~(\ref{eq:likelihood}), we are now in a position to test the hypothesis that these $N$ galaxies are drawn from some underlying model distribution.
To do so, we use the Monte-Carlo sampling code {\tt emcee} \citep{Foreman-Mackey+13} to randomly draw sets of $N$ galaxies from the weighted distribution function describing a given model.
Specifically, we generate sets of $N=51$ galaxies such that the $i^{\rm th}$ galaxy in this set (where $i=1,...,N$) is drawn from the weighted probability distribution function $\propto W(M, {\rm SFR}) \rho \left( M, {\rm SFR}, \vert z_i \right)$, where $z_i$ is the redshift of the $i^{\rm th}$ galaxy in the comparison FRB host-galaxy sample.
We repeat this process 32,000 times, effectively generating a large number of mock data-sets (each with $N=51$ galaxies) that can be compared to the observed data.
The choice of 32,000 mock data sets is arbitrary, but chosen to ensure sufficiently large statistics.

Using Equation~(\ref{eq:likelihood}), we calculate the likelihood $\mathcal{L}_{\rm mock}$ for each of these 32,000 mock data-sets. We then compare the likelihood of the real (FRB) data to the likelihoods of these mock data-sets.
If the FRB host-galaxy data is consistent with the model distribution, then its likelihood should be similar to the typical likelihood value of a mock data-set drawn from this distribution.
To estimate the likelihood of a ``typical'' mock data-set, we create a cumulative distribution $P( \mathcal{L}_{\rm mock} \leq \mathcal{L})$ of the likelihoods calculated for our 32,000 realizations.
We then define the p-value of the observed data for a particular model as
\begin{equation}
\label{eq:pval}
    p = P \left( \mathcal{L}_{\rm mock} \leq \mathcal{L}_{\rm data} \,\vert\, {\rm model} \right)
\end{equation}
where $\mathcal{L}_{\rm data}$ is the likelihood of the observed FRB host-galaxy sample within this model.
The p-value defined via Equation~(\ref{eq:pval}) can be interpreted as the probability that a set of $N$ galaxies randomly drawn from the assumed model distribution would have a likelihood as low, or lower than that of the observed galaxy sample.
For example, a p-value of $p=0.05$ implies that there is only a 5\% chance that galaxies drawn from the given model distribution would have a likelihood as extreme as the FRB host-galaxy sample.
In following standard conventions, we adopt $p<0.05$ as a threshold p-value below which we consider a given model to be ruled out by the data (in the sense that there would only be a $<5\%$ probability of observing the sample of known FRB host-galaxies if the model were correct).

Because there are uncertainties associated with inferred properties of the observed FRB host-galaxies (see Figure~\ref{fig:weighted_dists}), there is a corresponding uncertainty or spread in the likelihood value $\mathcal{L}_{\rm data}$ calculated for this sample.
To account for this fact, we use the error bars quoted in \cite{Gordon+23} and \cite{Sharma+24}
to generate a distribution of $\mathcal{L}_{\rm data}$.
Specifically, we use the 68\% confidence limits on stellar mass and SFR to define an asymmetric log-normal distribution (a distribution that has different standard deviations above and below the mean) for each of these quantities. We then repeatedly sample the $N=51$ host galaxies allowing the masses and SFRs of each galaxy to vary within these uncertainty limits. We then calculate the likelihood for each of these realizations and end up with a distribution of $\mathcal{L}_{\rm data}$ values.
We incorporate this into our final results by defining a 90\% upper limit on the p-value of a particular model given the observed data {\it and its variation}. This $p_{90}$ metric is found using the same procedure as described above, but comparing the mock data against the 90$^{\rm th}$ percentile value of $\mathcal{L}_{\rm data}$. 
To differentiate it from $p_{90}$, we refer to the nominal p-value obtained using the quoted (median) FRB host galaxy properties as $p_{\rm nom}$.

In comparison to $p_{\rm nom}$, $p_{90}$ is a more conservative limit. However, in practice we find that the two values are typically very close to one another (see Figure~\ref{fig:pvals}). In other words, accounting for uncertainties in the inferred stellar masses and SFRs of FRB hosts does not appreciably change the constraints on the models we have tested in this work.
As a caveat, we note that the quoted uncertainties on the inferred SFRs may be underestimated. We leave it to future work to investigate the potential impact of assuming larger uncertainties on these values.

Our overall procedure is illustrate in Figure~\ref{fig:pvals} for the mass-weighted, SFR-weighted, and GC-weighted models (panels a, b, and c, respectively). The solid black curve in each panel shows the cumulative distribution function (CDF) of the likelihoods of mock data-sets drawn from each of the three models, $P( \mathcal{L}_{\rm mock} \leq \mathcal{L})$. The vertical blue curves show the nominal likelihood calculated for the observed FRB host-galaxy sample $\mathcal{L}_{\rm data}$ under each of the three models (different panels).
The shaded vertical blue band surrounding each blue curve shows the range of variation of $\mathcal{L}_{\rm data}$, considering uncertainties in the derived FRB host-galaxy properties.
The value at the intersection of the vertical blue curve with the solid black curve defines the nominal p-value of the model given the observed data, $p_{\rm nom}$. This is marked with a red cross. The model is ruled out if the implied p-value falls below the horizontal dashed-grey curve that marks $p=0.05$.
Similarly, a more conservative p-value denoted $p_{90}$ is defined by the intersection of the upper-bound of the blue shaded region with the solid black curve.

\section{Results and implications}
\label{sec:results}

\begin{deluxetable}{lccccc}[]
\tablecaption{
The p-values of different models for FRB host-galaxies, as obtained by our methodology.
The `conservative' sub-sample considers only FRB host-galaxies that do not require any extension of the distribution (see \S\ref{sec:sample}).
\label{tab:pvals}}
\tablewidth{0pt} 
\tablehead{ & & \colhead{Mass$^{a}$} & \colhead{SFR$^{b}$} & \colhead{GC$^{c}$} & \colhead{SFR+Z$^{d}$}}
\tabletypesize{\small} 
\startdata 
    Full Sample & $p_{90}$$^{e}$ & 0.05 & 0.03 & 0.01 & 0\\
    ($N=51$) & $p_{\rm nom}$$^{f}$ & 0.04 & 0.03 & 0.01 & 0\\
    \hline
    Conservative & $p_{90}$ & 0.23 & 0.06 & 0.08 & \vline~~ 0.02$^{g}$~~~\\
    ($N=23$) & $p_{\rm nom}$ & 0.22 & 0.05 & 0.07 & \vline~~ 0.01$^{g}$~~~\\
\enddata 
\tablenotetext{}{$^{a}$Mass-weighted model (Equation~\ref{eq:weights_mass}); $^{b}$SFR-weighted model (Equation~\ref{eq:weights_SFR}); $^{c}$Globular-cluster-weighted model (Equation~\ref{eq:weights_GC}); $^{d}$SFR+Z model defined via Equation~(\ref{eq:weights_metallicity}) and following \cite{Sharma+24}; 
$^{e}$The 90\% upper confidence limit on the p-value considering the quoted uncertainties in inferred FRB host-galaxy masses and SFRs;
$^{f}$The nominal p-value obtained for the quoted (median) masses and SFRs of each FRB host galaxy.
$^{g}$Applied to the subsample of FRB host-galaxies with $z<0.2$ ($N=20$).
}
\end{deluxetable} 

Table~\ref{tab:pvals} summarizes the resulting p-values for the mass-weighted, SFR-weighted, and GC-weighted models.
Using our full sample of $N=51$ host galaxies, we can rule out 
the SFR and GC models at a significance of $p \approx 0.01$, $p \approx 0.03$ respectfully, while the mass-weighted model is only marginally ruled out ($p \approx 0.05$).
This implies that FRBs do not linearly track SFR in the Universe but are potentially consistent with following stellar mass, though only marginally so (as we discuss in \S\ref{sec:CDFs}, we disfavor this scenario). The results also show that FRBs, as a whole population, are inconsistent with tracking the mass of GCs.
Considering a conservative sub-sample of $N=23$ FRBs whose host galaxies do not require any extension of the \cite{Leja+22} distribution function (see \S\ref{sec:sample} and Tables~\ref{tab:mass_cutoff},\ref{tab:low_z}), we cannot rule out any of the three models, although the p-values of both SFR and GC models are only marginally consistent with the data (given our predetermined threshold of $p=0.05$).

\cite{Rao+25} have recently used GC simulations to show that NSs born through binary white-dwarf coalescence in GCs can contribute at most $\lesssim 1\%$ to the formation rate of FRB sources at $z \lesssim 1$.
Our results add to these findings using alternative methods, and support the conclusion that FRBs are inconsistent with a GC-origin as a whole population.
This may not be a very surprising conclusion, but it is important given the identification of at least one FRB to a GC \citep{Kirsten+22}. The relative fraction of FRB sources that could be located in GCs remains an open question.  
Our analysis uses host-galaxy information to show that it is statistically unlikely that all FRBs originate from GCs. But a more realistic model whereby only a fraction of FRBs are associated with GCs, would likely be consistent with the current host-galaxy data
(e.g., \citealt{Rao+25} find that $\lesssim 1\%$ of FRBs could arise from GCs).

Our conclusion regarding the SFR-weighted model is in line with recent work by \cite{Sharma+24} and provides further evidence that FRBs are not unbiased tracers of star-formation.
In their analysis, \cite{Sharma+24} used KS tests on the stellar-mass distributions to show that FRBs are inconsistent with linearly tracking SFR. In particular, they found a dearth of low-mass galaxies in the FRB host-galaxy sample in comparison to the expectation if FRB sources simply tracked SFR in the Universe.
This led \cite{Sharma+24} to suggest a modified model whereby FRBs track SFR, but only above a certain metallicity threshold.

To investigate this possibility we follow their work and define a SFR-weighted metallicity-dependent (`SFR+Z') model
using the following weighting-function 
\begin{equation}
\label{eq:weights_metallicity}
    \text{`SFR+Z':}~~~~
    W_i = {\rm SFR}_i 
    \times \left[ 1 + (M_i/M_Z)^{-\beta} \right]^{-1}
    ,
\end{equation}
where $M_Z$ is a cutoff mass 
(denoted $M_c$ in \citealt{Sharma+24})
that accounts for the fact that galaxies with lower mass (and hence, on average, lower metallicity) do not produce FRBs. The parameter $\beta$ controls the steepness of this cutoff. We use the values obtained by \cite{Sharma+24} for a subsample of FRB host galaxies at low redshifts, $z<0.2$: $\log_{10}(M_Z/M_\odot) = 8.98$, and $\beta \simeq 37,276$. 
For such large $\beta$, the sigmoid in the weighting-function (the second term in Equation~\ref{eq:weights_metallicity}) essentially becomes a step-function, selecting only galaxies with masses greater than $M_Z$. 
According to the mass-metallicity relations adopted in \cite{Sharma+24}, this threshold mass corresponds to a threshold oxygen abundance of $12 + \log({\rm O}/{\rm H}) \sim 8$ and a metallicity of $\log(Z/Z_\odot) \sim -0.6$.

Taken as-is, the SFR+Z model with the \cite{Sharma+24} parameters is completely ruled out if applied to the entire data-set of $N=51$ FRBs, with a p-value of $p \approx 0$. This is because one source in our sample, FRB~20210117A, has a stellar mass below $M_Z$. The model however predicts that there should be no FRBs below $M_Z$ given the extreme value of $\beta$ (the weight for such a galaxy would be $W_i \approx 0$). The likely reason this was not noticed by \cite{Sharma+24} is because that work focused only on FRB hosts at redshifts $z<0.2$ for this portion of the analysis. FRB~20210117A has a redshift of $z=0.2145$ (see Table~\ref{tab:mass_cutoff}) and therefore was not incorporated into the sub-sample that \cite{Sharma+24} investigated for the SFR+Z model (although this source is in their broader sample).

In principle the SFR+Z model of \cite{Sharma+24} could be extended by, for example, reanalyzing the model on FRB hosts at higher redshifts and updating the best-fit $M_Z$ and $\beta$ parameters, and/or by including redshift evolution of these parameters. However this is outside the scope of our current work.
Instead, we choose to check the validity of the \cite{Sharma+24} model as-is (without any modifications) by limiting the comparison to the domain in which the model was originally derived: for galaxies at redshifts $z<0.2$.

We run our analysis with the weighted-distribution defined by Equation~(\ref{eq:weights_metallicity}), and using only the sub-sample of $N=20$ FRB host galaxies that have $z<0.2$ (see Table~\ref{tab:low_z}). 
We find $p_{\rm nom}=0.01$ and $p_{90}=0.02$ for the nominal and 90\% upper-bound p-values, indicating that even for the subsample of $z<0.2$ galaxies, this model can be ruled out at high significance.
This is perhaps surprising given that 
we examine the same FRB host-galaxy subsample and that
our methodology is generally more conservative than previous approaches. We show in more detail in \S\ref{sec:CDFs} that the fundamental reason we are able to place such stringent constraints on this model is because our methodology uses the two-dimensional (bivariate) information content of galaxy's stellar-mass {\it and} SFR, while \cite{Sharma+24} only considered the distributions in stellar-mass.

\subsection{Mixed-Population Model}
\label{sec:mixed}

\begin{figure}
    \includegraphics[width=\columnwidth]{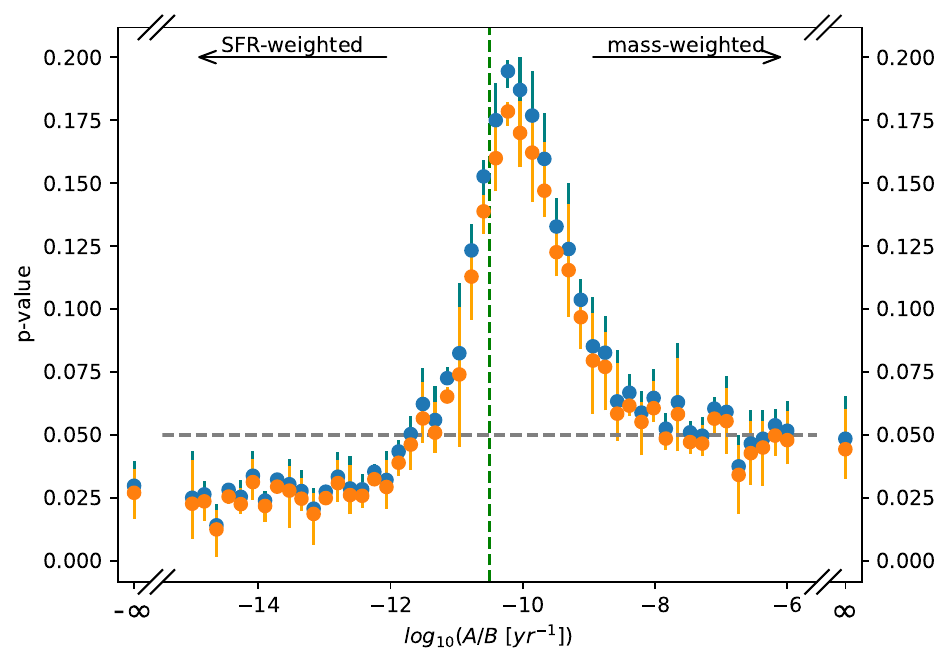}
    \caption{
    Consistency of `mixed' models which assume that FRB progenitors track a linear combination of stellar mass and star formation as parameterized via Equation~(\ref{eq:weights_AB}). The p-value of this model is shown as a function of $A/B$,
    the relative weight of each component.
    Orange points show the nominal p-value, while blue points show the more conservative $p_{90}$ (see \S\ref{sec:model_testing} and Figure~\ref{fig:pvals} for further details).
    As $\log_{10}(A/B) \to \infty$
    the mixed model reduces to the mass-weighted scenario. Conversely, the left end of the plot shows the limit 
    $\log_{10}(A/B) \to -\infty$ which describes the SFR-weighted scenario. 
    Points that fall below the dashed-grey curve have $p<0.05$ and are considered ruled out at this threshold.
    The peak near the center of the plot indicates that a mixed model is most consistent with the data. 
    Our preferred model has $A/B = 10^{-10.5}\,{\rm yr}^{-1} \approx 0.03\,{\rm Gyr}^{-1}$ and is marked with a dashed-green vertical curve.
    }
    \label{fig:ab_frac}
\end{figure}

We continue by investigating a mixed model in which FRBs track a linear combination of both stellar mass and SFR, as described via Equation~(\ref{eq:weights_AB}).
Figure~\ref{fig:ab_frac} shows the resulting p-values for such a model, as a function of $A/B$---the ratio of the stellar mass prefactor and the SFR prefactor in Equation~(\ref{eq:weights_AB}).
Orange (blue) points show $p_{\rm nom}$ ($p_{90}$) for different values of $A/B$. The right-hand side of this plot corresponds to the purely mass-weighted model, while the left-hand side corresponds to the SFR-weighted scenario. The p-values in either case are consistent with the values quoted in Table~\ref{tab:pvals} for those two models. 
The SFR-weighted end falls below our threshold $p=0.05$ value (horizontal dashed-grey curve) implying that this models is ruled out, as previously discussed.
The mass-weighted end asymptotes to a value slightly above $p=0.05$ and can cannot be strictly ruled out by this analysis, however there is only marginal support for such a model (see also \S\ref{sec:CDFs}).
Alternatively, Figure~\ref{fig:ab_frac} shows that a mixed model with  $A/B \sim 10^{-10}\,{\rm yr}^{-1}$ seems to be preferred by the current data.
This suggests that FRBs could track a combination of both stellar mass and star formation. 

This is interesting, as a similar two-component model has been invoked to explain the diversity of Type Ia SNe hosts \citep[e.g.,][]{Scannapieco+05}. 
Indeed, the possible similarity between FRB and Type Ia SNe progenitors has already been pointed out in the literature \citep{Margalit+19,Eftekhari+25}.
At the risk of over interpreting Figure~\ref{fig:ab_frac}, we examine more quantitative implications of this result.

To be concrete, we focus on a fiducial mixed model with $A/B = 10^{-10.5} \, {\rm yr}^{-1}$, shown as a dashed-green vertical curve if Figure~\ref{fig:ab_frac}. Such a model is supported by our primary analysis, and has a p-value of $p_{90} \simeq 0.19$ ($p_{\rm nom} \simeq 0.15$). 
Given the wide range of $A/B$ values that are consistent with high p-values (see Figure~\ref{fig:ab_frac}), this particular choice of is somewhat arbitrary. We focus on this `preferred' value because we find that, in addition to falling within the peak of the $A/B$ p-values seen in Figure~\ref{fig:ab_frac}, it also reasonably fits the marginalized stellar-mass and SFR CDFs, as discussed in \S\ref{sec:CDFs} (see also Figure~\ref{fig:CDFs}). Higher values of $A/B$ that place more weight on stellar-mass, while supported by the p-values shown in Figure~\ref{fig:ab_frac}, would provide poorer fits to the marginalized stellar-mass CDFs. We therefore consider our fiducial $A/B = 10^{-10.5} \, {\rm yr}^{-1}$ as being preferred over, say $A/B \sim 10^{-9} \, {\rm yr}^{-1}$, although we cannot strictly rule out such $A/B$ values with our analysis.

The preferred value of
$A/B \approx 3 \times 10^{-11} \, {\rm yr}^{-1}$ that we find here 
is close to the 
value inferred for Type Ia SNe, $A/B \approx 1.7^{+2.3}_{-0.9} \times 10^{-11} \, {\rm yr}^{-1}$ \citep{Scannapieco+05}.
This implies that the progenitors of FRBs and Type Ia SNe may be related. 
In particular, this aligns with the idea that FRBs might be associated with accretion-induced collapse of a white-dwarf, as such events are thought to form through similar channels as Type Ia SNe \citep{Kashiyama&Murase17,Waxman17,Margalit+19,Eftekhari+25}.
We note however that there are significant uncertainties associated with our preferred $A/B$ value, and therefore caution is warranted in interpreting these results. As mentioned in \S\ref{sec:methods}, using a different SED modeling code can potentially affect the specific star-formation rate of galaxies ${\rm sSFR} \equiv {\rm SFR}/M$, and hence also the inferred A/B values, by up to an order of magnitude \citep[e.g.,][]{Leja+20}. 
Although the A/B values found by \cite{Scannapieco+05} are not obtained using {\tt Prospector}, their methodology differs substantially from our present work, and does not rely on detailed galaxy SED fitting. Instead, \cite{Scannapieco+05} use a sub-sample of host galaxies that are categorized as either ``star-forming'' or ``quiescent'', and determine the A and B values by calibrating to the observed CCSN rate in these galaxies.
Furthermore, recent work by \cite{Nugent+25} re-analyzed Type Ia SNe host galaxies uniformly using {\tt Prospector}, and found that the A/B values inferred by \cite{Scannapieco+05} are consistent with those inferred using this SED-fitting code. 

Proceeding with these caveats in mind, we consider some implications of the preferred value of $A/B \approx 10^{-10.5}\,{\rm yr}^{-1}$ that we find for FRBs.
One interpretation of the mixed model is that there are
two distinct formation channels (components) for FRB sources: one that follows star-formation and one that is dependent on stellar mass.
Viewed this way, the parameter $A/B$ can be interpreted as a 
${\rm sSFR}$
that delineates between the two components. FRB sources located in host-galaxies whose specific SFR exceeds
${\rm sSFR} > A/B \approx 10^{-10.5}\,{\rm yr}^{-1} \approx 0.03 \, {\rm Gyr}^{-1}$
are likely to have formed through the star-formation channel, while sources in galaxies with ${\rm sSFR} < A/B$ are more likely to have formed via the mass-dependent channel.
Another interpretation of the mixed model is that FRB sources are born through a single channel, but one that has an extended time-delay distribution (meaning, a fraction of sources are formed long after their progenitor stars were born). Although the model does not explicitly account for the delay-time distribution, the combination of stellar-mass and SFR can mimic this behavior. 
This is the common interpretation for short GRBs and Type Ia SNe \citep[e.g.,][]{Scannapieco+05,Fong+13}. The extended delay time in these settings is then typically associated with the gravitational-inspiral time of a compact object binary.

Motivated by the similarity with Type Ia SNe host-galaxies, 
we follow the analysis of \cite{Scannapieco+05} and adopt a toy-model for galaxy formation whereby stellar mass grows only as a consequence of star-formation such that $M(t) = \int {\rm SFR}(t) dt$ and taking an exponential star-formation history, ${\rm SFR}(t) = {\rm SFR}_0 e^{-t/\tau}$ with $\tau = 2\,{\rm Gyr}$ (this neglects galaxy mergers and hierarchical structure formation).

Using this simplified approach, we investigate implications of our preferred mixed-population model. We assume that FRBs track a combination of both stellar mass and star-formation, with $A/B = 10^{-10.5}\,{\rm yr}^{-1}$ (Equation~\ref{eq:weights_AB}). The birth-rate of FRB sources originating from the stellar-mass component at some time $t$, $\left. \mathcal{R}_{\rm FRB}(t) \right\vert_{A}$, can be compared to the rate of sources tracking the SFR component, $\left. \mathcal{R}_{\rm FRB}(t) \right\vert_{B}$, 
where $\mathcal{R}_{\rm FRB} = \left. \mathcal{R}_{\rm FRB} \right\vert_{A} + \left. \mathcal{R}_{\rm FRB} \right\vert_{B}$ is the total formation rate of FRB sources. 
The relative rate in this scenario is

\begin{equation}
    \frac{
    \left. \mathcal{R}_{\rm FRB}(t) \right\vert_{A}
    }{\left. \mathcal{R}_{\rm FRB}(t) \right\vert_{B}} = \frac{A}{B} \tau \left( e^{t/\tau} - 1 \right) \underset{t=10\,{\rm Gyr}}{\sim} 
    9.
\end{equation}
That is, after $t=10\,{\rm Gyr}$, only $\sim$10\% (one out of 10) of FRB sources are born via channels that track SFR.
Conversely, the ratio of the total number of FRB sources born through either channel throughout the lifetime of such a galaxy is
\begin{equation}
    \frac{\left. N_{\rm FRB}(t) \right\vert_{A}}{\left. N_{\rm FRB}(t) \right\vert_{B}} = \frac{A}{B} \tau \frac{(t/\tau)e^{t/\tau} + 1}{e^{t/\tau} - 1} \underset{t=10\,{\rm Gyr}}{\sim} 
    0.3.
\end{equation}
This implies that over the lifetime of a galaxy roughly 
$\sim$80\% (one out of 1.3) 
of FRB sources will be born through SFR-related channels.
Of course, it is important to bear in mind the order-of-magnitude uncertainty in the allowed $A/B$ values, as well as the grossly over-simplified galaxy growth model that goes into these estimates. They are nonetheless illustrative and useful as a comparison to Type Ia SNe progenitors.

Using our preferred mixed model we can also predict the fraction of quiescent FRB host galaxies. We simulate a mock population of galaxies drawn from the weighted distribution defined by Equation~(\ref{eq:weights_AB}) with $A/B=10^{-10.5}\,{\rm yr}^{-1}$ at redshift $z=0.3$. We then classify galaxies as quiescent, transitioning, or star-forming following the methodology described in \cite{Gordon+23}.
Specifically, we use the mass-doubling number $\mathcal{D} = {\rm sSFR} \times t_{\rm H}(z)$, 
where $t_{\rm H}(z)$ is the Hubble time at redshift $z$,
and classify galaxies as quiescent if $\mathcal{D}<1/20$, transitioning if $1/20<\mathcal{D}<1/3$, and star-forming if $\mathcal{D}>1/3$ \citep{Tacchella+22}.
According to this criterion we find that 13\% of the simulated galaxies are quiescent, 10\% are transitioning, and 77\% are star-forming at $z=0.3$.
At redshift $z=0.5$ the associated breakdown is 10\% quiescent, 8\% transitioning, and 82\% star-forming.

From the quoted results in \cite{Gordon+23}, \cite{Bhardwaj+24}, and \cite{Sharma+24}, the breakdown of the currently-adopted sample of FRB hosts is: 3/51 (6\%) quiescent galaxies, 9/51 (18\%) transitioning, and 39/51 (76\%) star-forming.
This is broadly in line with the expected breakdown for our preferred mixed model, although quiescent galaxies are somewhat underrepresented and transitioning galaxies overrepresented in the current FRB sample. We therefore predict that more quiescent FRB hosts will be found as the sample of localized FRBs increases. In fact, \cite{Shah+25} and \cite{Eftekhari+25} recently reported the discovery of an FRB associated with a quiescent elliptical galaxy, in line with these predictions.

We note again that our preferred A/B value can be interpreted as a sSFR. The sSFR of a ``typical'' galaxy can be related to the inverse of the Hubble time, since the time-averaged SFR over the history of the galaxy roughly satisfies $\langle {\rm SFR} \rangle_t \times t_{\rm H} \sim M$. Our preferred A/B~$\sim 10^{-10}\,{\rm yr}^{-1}$ value is therefore consistent with the average sSFR expected for a ``typical'' low-redshift galaxy, at least at the order-of-magnitude level.
This can be interpreted as a consequence of FRB sources being found in a wide range of galaxy types. If all FRBs were instead formed in extreme starburst galaxies (or alternatively, in highly-quiescent galaxies), their sSFRs would be much higher (lower) than $t_{\rm H}^{-1}$ and the inferred A/B values would differ markedly from the inverse Hubble time.

As we were completing our work, a paper by \cite{Loudas+25} came out which conducted a similar study and also examined a mixed model. The formalism of the mixed model in \cite{Loudas+25} differs from that of our present work. Instead of assigning different weights $A$ and $B$ to the mass- and SFR-weighted components, \cite{Loudas+25} construct mock samples in which a fraction $\chi_{\rm SFR}$ of the samples are drawn from a SFR-weighted distribution while the remaining fraction $1-\chi_{\rm SFR}$ are drawn from a mass-weighted distribution. 
These two models are conceptually similar but physically distinct.
$\chi_{\rm SFR}$ is a property of a sample or population, while our weighting function~(\ref{eq:weights_AB}) is a property of individual galaxies.
Nevertheless, with some assumptions about the population of galaxies under consideration, the two parameterizations can be related to one another.
This conversion is redshift dependent, so
a given $A/B$ will correspond to different $\chi_{\rm SFR}$ values depending on the redshift, as we show below.

For a sample of galaxies drawn at a single redshift,
our $A/B$ parameter translates into
\begin{equation}
    \chi_{\rm SFR} = \left[ 1 + \frac{A/B}{N(z)} \right]^{-1}
    ,
\end{equation}
where
\begin{equation}
    N(z) = \frac{
    \int {\rm SFR} \rho\left(M,{\rm SFR}|z\right) \,d{\rm SFR} \,dM
    }{
    \int M \rho\left(M,{\rm SFR}|z\right) \,d{\rm SFR} \,dM
    }
\end{equation}
is a redshift-specific normalization that can be interpreted as the ratio between the cosmic star-formation density $\Psi(z)$ 
to the cosmic stellar-mass density $\rho_*(z)$ at redshift $z$ .
Within this framework our preferred value of $A/B = 10^{-10.5}\,{\rm yr}^{-1}$ translates into $\chi_{\rm SFR} \approx 0.65$ at redshift $z=0.1$, $\chi_{\rm SFR} \approx 0.74$ at $z=0.3$, and $\chi_{\rm SFR} \approx 0.80$ at $z=0.5$.
These values are all consistent with the constraints on $\chi_{\rm SFR}$ obtained by \cite{Loudas+25}.

One major difference between the conclusions of our present work and \cite{Loudas+25} is our finding that a mixed model is preferred over a SFR-weighted scenario (which we explicitly rule out). In comparison, \cite{Loudas+25} place lower bounds on $\chi_{\rm SFR}$ but prefer a purely SFR-weighted model. The reason we arrive at different conclusions is because our methodology considers the bivariate distribution in both stellar mass and SFR, whereas \cite{Loudas+25} relied on the stellar-mass distributions in their primary analysis. This again highlights the importance of using the multivariate nature of host-galaxy data.

\subsection{Marginalized Distribution Functions}
\label{sec:CDFs}

\begin{figure*}
    \begin{subfigure}[b]{0.9\textwidth}
        \centering
        \includegraphics[width=\textwidth]{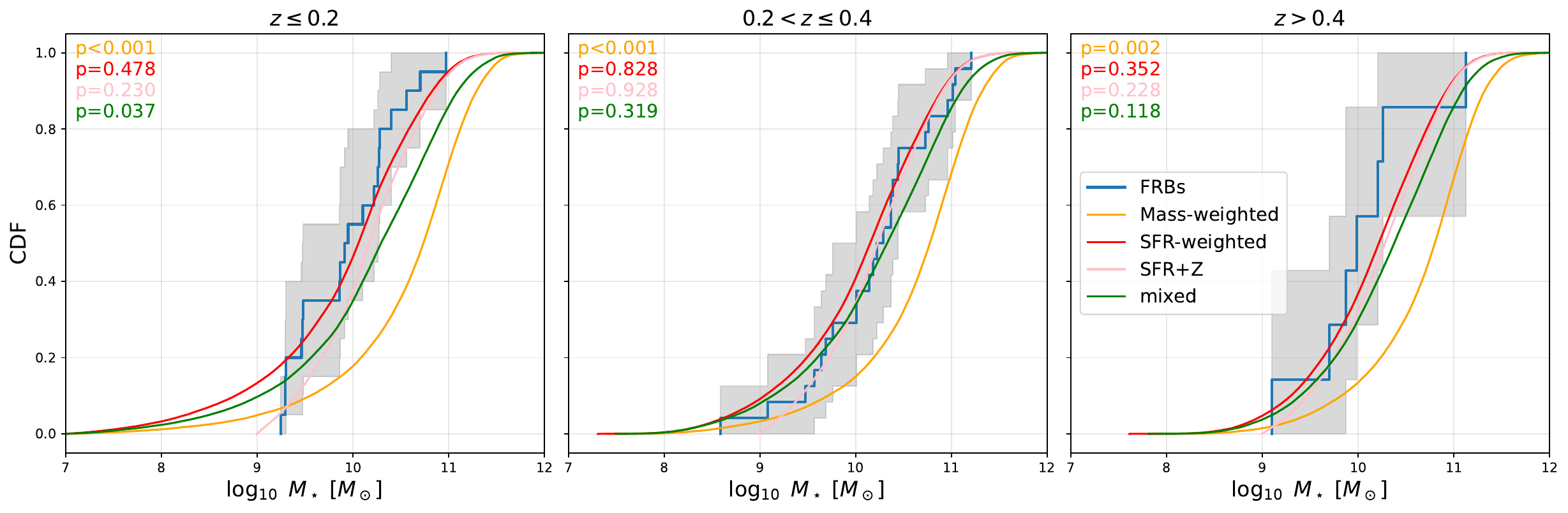}
    \end{subfigure}
    \hfill
    \begin{subfigure}[b]{0.9\textwidth}
        \centering
        \includegraphics[width=\textwidth]{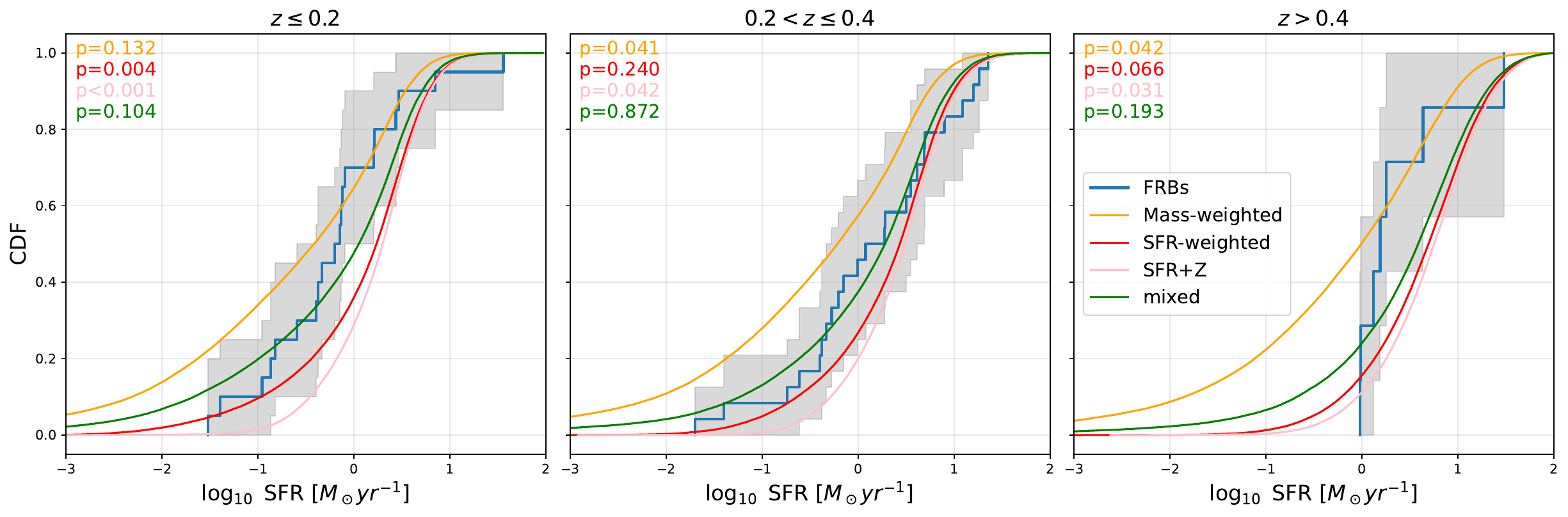}
    \end{subfigure}
    
    \caption{
    Marginalized stellar-mass (top panels) and SFR (bottom) distribution functions, split into three redshift bins.
    Blue curves show the CDFs of the FRB hosts within each redshift bin, and shaded grey areas show the 95\% CI on these CDFs obtained by bootsrapping.
    The colored curves in each panel show different models discussed in the text: the mass-weighted model (yellow), the SFR-weighted model (bright red), the SFR+Z model proposed by \cite{Sharma+24} (pink), and our preferred mixed model for which $A/B = 10^{-10.5}\,{\rm yr}^{-1}$ (green).
    The p-values resulting from KS tests of each model against the FRB data are listed in the top-left corner of each plot.
    See \S\ref{sec:CDFs} for further details.
    The model samples in each subplot are generated at a redshift corresponding to the median redshift of FRB host galaxies within that bin. 
    }
    \label{fig:CDFs}
\end{figure*}

To further investigate the results discussed above, we plot in Figure~\ref{fig:CDFs} the univariate marginalized CDFs as a function of either stellar-mass (top panels) or SFR (bottom panels). To limit the effects of redshift evolution we separate these CDFs into three redshift bins selected as: $z<0.2$, $0.2 < z < 0.4$, and $z > 0.4$. These particular redshift bins were chosen to align with those in \cite{Sharma+24} and \cite{Loudas+25} in order to better facilitate comparison with these works.
In each panel we plot the CDFs of the mass-weighted model (yellow), the SFR-weighted model (bright red), the \cite{Sharma+24} SFR+Z model (dark red), and our preferred mixed model (green) with $A/B = 10^{-10.5}\,{\rm yr}^{-1}$.
We also plot the FRB host-galaxy data in blue along with the associated 95\% confidence interval (CI) obtained by bootstrapping, using the following procedure. For the given dataset of size $N=51$ listing the FRB host-galaxy properties (either stellar-mass or SFR), we create a bootsrapped dataset by sampling the original dataset $N$ times, with repetitions allowed. We generate 1,000 bootstrapped samples in this way, and calculate the 95\% CI of the resulting CDFs.
Figure~\ref{fig:CDFs} also lists the p-values resulting from a KS test of each model against the FRB data, quoted in the top-left corner of each plot and following the color-coding convention for the models.

Figure~\ref{fig:CDFs} helps illustrate the strengths and weaknesses of different models, and why considering the full bivariate distribution in both stellar-mass and SFR is critical.
For example, when examined only as a function of stellar-mass, the SFR+Z model describes the FRB data reasonably well. However, when instead viewed as a function of SFR, the SFR+Z model poorly matches the data. Correspondingly a KS test of this model against the FRB host-galaxy SFR distribution receives low p-values. This helps explain why we rule out the best-fit SFR+Z model promoted by \cite{Sharma+24} despite our using the exact same dataset. As previously mentioned, \cite{Sharma+24} developed this model based on an analysis that used only information about the stellar-mass distributions. Through that limited lens, the SFR+Z model indeed works quite well. However, when also accounting for information about the SFRs of galaxies this is no longer the case.
This is broadly true also of the simpler SFR-weighted model (bright red curves in Figure~\ref{fig:CDFs}). The fact that such a model over-predicts the SFRs of FRB host-galaxies can also be seen by eye in Figure~\ref{fig:weighted_dists} (middle panel).

Similarly, Figure~\ref{fig:CDFs} shows that the mass-weighted model systematically over-predicts the stellar-mass compared to the observed FRB host galaxies. This is not surprising, given that the mass-weighted model is marginally ruled out by our analysis (it has a p-value of $\approx 0.05$; see Table~\ref{tab:pvals}).
As previously discussed, the methodology of our primary analysis is conservative in the sense that models that achieve a high p-value may not necessarily be representative of the underlying distribution of the data, however models that garner low p-values can robustly be ruled out. The fact that the mass-weighted model does not appear to fit the marginalized CDFs is therefore not in tension with its moderate p-value.

In contrast to the mass-weighted or SFR-weighted (or SFR+Z) models, Figure~\ref{fig:CDFs} shows that a mixed model with $A/B \sim 10^{-10.5}\,{\rm yr}^{-1}$ fits both the stellar-mass and SFR distributions of FRBs reasonably well, albeit with some deficiencies in the lowest redshift bin. This provides further evidence in favor of such a model, as already indicated by Figure~\ref{fig:ab_frac} (see discussion in \S\ref{sec:mixed}).
The combination of these results lead us to argue that a mixed model with these parameters is preferred over the other models considered here, and is one of the main conclusions of our present work.


Finally, we stress that the marginalized CDFs in-and-of-themselves would not, in our view, constitute sufficient evidence for such a claim. This is because such examinations are subject to the numerous issues discussed in \S\ref{sec:past_work}.
For example, the choice of stellar-mass and SFR as two independent axes is arbitrary from a data-driven point of view. Different linear combinations of $M$ and SFR would result in different CDFs, which could potentially lead to different conclusions. Furthermore, it is necessary to bin in redshift when creating such plots. This subjects the analysis to binning errors, and in particular the possibility that redshift-evolution within each bin or that redshift-dependent observational systematics may bias the results. We therefore view the analysis presented in Figure~\ref{fig:CDFs} above primarily as a consistency check, and as a means to help interpret and compare our results to previous work.

\section{Discussion and Summary}
\label{sec:discussion}

In this work we have presented a new framework for statistically testing models that describe the host-galaxy distributions of astrophysical sources. Our framework has the advantage of being applicable to sources that span a wide range in redshift, and accounts for redshift evolution in a way that is unbiased and model-agnostic, avoiding possible pitfalls to other approaches in the literature (such as binning or `transposition'; see \S\ref{sec:past_work}). 
In particular,
our method avoids redshift-dependent selection effects
by using the conditional probability $\rho\left( M, {\rm SFR} \,\vert z \right)$
and constructing redshift-matched mock catalogs (whereby each mock source is matched to the redshift of a real source in the astrophysical sample). 
Additionally, our methodology uses the multi-variate nature of the data in a self-consistent way---leveraging both host-galaxy stellar-mass and SFR measurements in a joint analysis, rather than examining each (or only one) property independently.

The primary focus of our current work has been to use this new framework to study the host galaxies of FRBs. Applying our method to a sample of $N=51$ FRB host-galaxies (see \S\ref{sec:sample}), we are able to rule out several previously proposed models and show that FRBs are instead consistent with tracking a combination of both star-formation and stellar-mass. 

Our main findings can be summarized as follows:
\begin{itemize}
    \item We have developed a novel method for comparing the host-galaxy properties of astronomical transients against models. Our methodology is designed to handle transients that span a wide range in redshift, and avoid pitfalls associated with redshift evolution of the underlying distribution. Importantly, it incorporates the bivariate information content of samples, accounting for both stellar masses and SFRs of galaxies in a self-consistent way.
    The generality of this method avoids various issues that can bias more-commonly-used univariate KS tests.
    Our approach is conservative: a model that garners a high p-value using our method may not necessarily well-describe the data, however a model that attains a low p-value can robustly be ruled out.
    
    \item Using our method we rule out the hypothesis that FRBs track SFR. 
    We can only marginally rule out the hypothesis that FRBs track stellar-mass, but we consider this scenario unlikely given the marginal p-values and the poor match between the resulting CDFs and the FRB data (Table~\ref{tab:pvals}; Figure~\ref{fig:CDFs}). 

    \item We find evidence that the sample of FRB host-galaxies is best-described by a `mixed' model where FRBs track a linear combination of both stellar-mass and SFR. 
    In particular, we find that the data is well-described by a model with $A/B \sim 10^{-10.5}\,{\rm yr}^{-1} \approx 0.03 \, {\rm Gyr}^{-1}$, where $A/B$ is the relative weighting of stellar-mass compared to SFR (Equation~\ref{eq:weights_AB}). This is supported both by the high p-value of such a model and by the agreement of the resulting marginalized CDFs (Figures~\ref{fig:ab_frac},\ref{fig:CDFs}).
    
    \item
    This preferred model is similar to inferences for Type Ia SNe, where an identical model with similar $A/B$ values works well \citep{Scannapieco+05}.
    This suggests 
    that FRB progenitors and Type Ia SNe may form through similar channels,
    potentially implicating accretion-induced collapse as an FRB formation pathway \citep[e.g.,][]{Margalit+19,Kremer+21,Lu+22,Eftekhari+25}.

    \item A prediction of this model is that the fraction of quiescent, transitioning, and star-forming host galaxies should be approximately 10\%, 10\%, and 80\%, respectively (see \S\ref{sec:mixed}). 
    This is broadly consistent with the current sample of FRB hosts, although quiescent galaxies are underrepresented in the current sample. We therefore predict that more quiescent FRB hosts will be found. This is consistent with the recent discovery of FRB~20240209A (\citealt{Shah+25,Eftekhari+25}; which is not included in our data set).
    Another prediction is that the fraction of quiescent galaxies should decrease at higher redshifts. 
    
    \item We examined the recently suggested metallicity-dependent SFR-weighted (`SFR+Z') model suggested by \cite{Sharma+24}, and rule out this model with high significance (Table~\ref{tab:pvals}). A primary reason we arrive at different conclusions than \cite{Sharma+24} is our examination of the full bivariate $M$-SFR distributions instead of only marginalized stellar-mass distributions. 
    We note that we only rule out the specific model presented in \cite{Sharma+24}. We have not examined a broader class of metallicity-dependent models, and it remains possible that there exist such models that are consistent with the data.

    \item We also test the hypothesis that all FRBs are produced in GCs. Based on host-galaxy demographics alone, we rule out this hypothesis at the $p=0.01$ level. 
    Of course, this does not preclude the possibility that only a subset of FRBs originate from GCs, consistent with at least one source 
    (\citealt{Kirsten+22}; see also \citealt{Rao+25}).
\end{itemize}

Our finding that the host-galaxy distribution of FRBs is best modeled as a mixed distribution that follows both stellar mass and star formation can be interpreted in two ways: (i) that there are two distinct formation channels for FRB sources, one that tracks star formation and another that tracks stellar mass, or (ii) that FRB sources are born through a single channel that has an extended delay-time distribution.
We suggest that the first interpretation can be tested by comparing the properties of FRBs that occur in galaxies whose ${\rm sSFR} > A/B \approx 0.03\,{\rm Gyr}^{-1}$ with FRBs in galaxies where ${\rm sSFR} < A/B$. 
This separates FRBs into two classes based on whether the source was more likely to have formed though the SFR- or stellar-mass-associated channels.
If such sources are truly distinct, one might expect their radio properties to differ from one another (although this must not necessarily be the case).
The second interpretation of our mixed model is consistent with the standard view of Type Ia SNe and short GRBs which are thought to form via channels that have extended time-delay distributions, such as compact object mergers.
Indeed, our results are consistent with previously suggested similarities between the host-galaxies of FRBs, Type Ia SNe and short GRBs \citep[e.g.,][]{Margalit+19,Eftekhari+25, Chen+24}.
Our conclusion---derived from FRB host galaxies---that FRB sources track a combination of both SFR and stellar-mass supports recent suggestions based on modeling the FRB redshift evolution and luminosity function (\citealt{Mo+25,Gupta+25}; although see \citealt{James+22,Shin+23} who, using similar methods, instead favor a scenario in which FRBs track SFR, or \citealt{Hashimoto+22,Chen+24} who conversely find that FRBs more-closely track stellar-mass).

It is important to note that there are various caveats to our analysis, and our findings should be interpreted with caution. In particular, there is a risk that the similarity in the inferred $A/B$ values for Type Ia SNe and FRBs is driven by a confounder, rather than reflecting an intrinsic similarity between the progenitors of these distinct transients. 
We note however that the possible connection between FRBs and Type Ia SNe is separately supported by direct comparison of FRB host-galaxy properties with Type Ia SNe hosts. A recent comparison of this type can be found in \cite{Sharma+24}.
Extended Data Figure~8 and Supplementary Table~5 in \cite{Sharma+24} show that the stellar-mass and SFR CDFs of Type Ia SNe hosts agree well with the current sample of FRB hosts. Similarly, Extended Data Figure~7 in that work shows that the offsets of Type Ia from their host galaxies agree well with FRB offsets.
These facts further support our current findings, obtained using a completely different methodology and without directly relying on Type Ia SNe data.

We encourage future work using our formalism to further investigate FRB progenitor models, especially as the sample of localized FRBs increases in size.
For example, models that depend non-linearly on mass or SFR, or a broader class of metallicity-dependent models could be investigated.
Furthermore, a more detailed treatment of the optical selection function,
as recently proposed by \cite{Loudas+25},
could be incorporated into our analysis. 
Finally, we note that our current framework requires extending the empirical galaxy distribution function at low redshifts $z \lesssim 0.2$. Such an extension was also necessary in the recent works by \cite{Sharma+24} and \cite{Loudas+25}. Using this method, the low-redshift extension extrapolates fitting formulae to the stellar-mass function and star-forming sequence obtained at higher redshifts. 
A different approach, which we have tested as well, is to assume that the galaxy distribution evolves very little between $z=0.2$ and $z=0$, and therefore that the distribution function is constant at these redshifts. This approach yields very similar results, and does not alter our primary conclusions. Both choices have drawbacks, and the only self-consistent approach would be to train a model on field-galaxy data that goes down to $z=0$. In our current work we choose a redshift extension that is consistent with previous works \citep{Sharma+24,Loudas+25}. Although this method introduces some model dependence, it is more realistic than assuming a no redshift evolution \citep[e.g.,][]{MadauDickinson14}.

In the present work we have also ignored the difference between repeating and apparently non-repeating FRBs. Future work could use our methodology to investigate whether these two classes of FRB sources may be associated with different progenitors based on each sub-sample's host galaxy properties (to the extent that they are truly separate classes).
Using different methods, \cite{Gordon+23} investigated this issue and found no statistically-meaningful difference between the hosts of repeating and apparently non-repeating FRBs (see also \citealt{Sharma+24}).

We also note that unmodeled radio selection effects could bias our results. For example, our findings imply that FRBs track a combination of both star-formation and stellar mass. If FRB sources born through either channel systematically differ in the properties of the radio bursts that they produce then our FRB host-galaxy sample may over- or under-represent one component over the other. This would presumably affect our inferred value of $A/B$ which represents the relative contribution of mass and SFR to the creation of FRB sources.

The source code used in our study is publicly available\footnote{\url{https://github.com/hoasaf3/host-galaxies-stats}} \citep{host_galaxies_stats}, and is designed to be applicable to broader contexts beyond the specific scope of FRB host-galaxies.
We caution that it is important that application of this framework be limited to galaxies whose photometry and/or spectroscopy have been modeled using the non-parametric SED-modeling code {\tt Prospector} \citep{Leja+17,Johnson+21}. 
This is important because our formalism uses a galaxy distribution that is trained on galaxies that have been modeled using this code, and recent work has shown that the choice of SED-modeling code can systematically affect the inferred galaxy properties \citep{Leja+19,Leja+22}.

\begin{acknowledgments}
We thank Alexa Gordon for helpful conversations and for commenting on an early version of this work, and Nick Loudas for insightful comments and for helping facilitate a comparison of our respective works.
We also thank the anonymous referee for comments and suggestions that helped improve this manuscript.
\end{acknowledgments}

\textit{Author's Note (AH):} This paper has taken many turns since its inception, but its earliest draft bore the title ``Can Babies Be Born in a Nursing Home?''—a nod to the question that drove the initial version of this work: how likely is it for an FRB, widely considered to originate from a young magnetar, to come from a GC—an environment dominated by old stars—as in the case of FRB~20200120E. The scope has since broadened, but I still wanted to give that title a small place in the final version.

\software{
        {\tt numpy} \citep{numpy}
        {\tt matplotlib} \citep{matplotlib},
        {\tt emcee} \citep{Foreman-Mackey+13},
        {\tt astropy} \citep{astropy+13, astropy+18, astropy+22},
        {\tt scipy} \citep{scipy},
        {\tt pandas} \citep{pandas},
        {\tt statsmodels} \citep{statsmodels}
          }


\appendix

This Appendix contains Tables listing several properties relevant to the FRB host-galaxy sample used in this work.

\begin{table}
    \centering
    \caption{FRB host galaxies that fall below the threshold mass $M_{\rm th}(z)$}
    \label{tab:mass_cutoff}
    \begin{tabular}{cccc} 
        \hline
		FRB & $z$ & $\log_{10} \left( M / M_\odot \right)$ & $\log_{10} \left[ M_{\rm th}(z) / M_\odot \right]$\\
		\hline
            20190711A & 0.5218 & 9.10 & 9.16 \\
            20210117A & 0.2145 & 8.59 &  8.65 \\
        \hline
    \end{tabular}
\end{table}

\begin{table}
    \centering
    \caption{List of the 27 FRB host galaxies with redshift $z<0.25$.}
    \label{tab:low_z}
    \begin{tabular}{cc} 
        \hline
		FRB & $z$ \\
		\hline
            20220319D  & 0.01120 \\
            20231120A  & 0.03680 \\
            20220207C  & 0.04330 \\
            20220509G  & 0.08940 \\
            20230124A  & 0.09390 \\
            20220914A  & 0.11390 \\
            20230628A  & 0.12700 \\
            20220920A  & 0.15820 \\
            20221101B   & 0.23950 \\
            20220825A  & 0.24140 \\
            20220307B   & 0.24810 \\
            20180916B  & 0.03300 \\
            20190520B  & 0.24170 \\
            20190608B  & 0.11780 \\
            20190714A  & 0.23650 \\
            20191001A  & 0.23420 \\
            20200430A  & 0.16070 \\
            20201124A  & 0.09800 \\
            20210117A  & 0.21450 \\
            20210410D  & 0.14150 \\
            20210807D  & 0.12930 \\
            20211127I  & 0.04690 \\
            20211212A  & 0.07070 \\
            20181220A  & 0.02746 \\
            20181223C  & 0.03024 \\
            20190418A  & 0.07132 \\
            20190425A  & 0.03122 \\
        \hline
    \end{tabular}
\end{table}


\bibliography{references}{}

\begin{thebibliography}{}
\expandafter\ifx\csname natexlab\endcsname\relax\def\natexlab#1{#1}\fi
\providecommand{\url}[1]{\href{#1}{#1}}
\providecommand{\dodoi}[1]{doi:~\href{http://doi.org/#1}{\nolinkurl{#1}}}
\providecommand{\doeprint}[1]{\href{http://ascl.net/#1}{\nolinkurl{http://ascl.net/#1}}}
\providecommand{\doarXiv}[1]{\href{https://arxiv.org/abs/#1}{\nolinkurl{https://arxiv.org/abs/#1}}}

\bibitem[{ {Astropy Collaboration} {et~al.}(2013){Astropy Collaboration}, {Robitaille}, {Tollerud}, {Greenfield}, {Droettboom}, {Bray}, {Aldcroft}, {Davis}, {Ginsburg}, {Price-Whelan}, {Kerzendorf}, {Conley}, {Crighton}, {Barbary}, {Muna}, {Ferguson}, {Grollier}, {Parikh}, {Nair}, {Unther}, {Deil}, {Woillez}, {Conseil}, {Kramer}, {Turner}, {Singer}, {Fox}, {Weaver}, {Zabalza}, {Edwards}, {Azalee Bostroem}, {Burke}, {Casey}, {Crawford}, {Dencheva}, {Ely}, {Jenness}, {Labrie}, {Lim}, {Pierfederici}, {Pontzen}, {Ptak}, {Refsdal}, {Servillat}, \& {Streicher}}]{astropy+13}
{Astropy Collaboration}, {Robitaille}, T.~P., {Tollerud}, E.~J., {et~al.} 2013, \bibinfo{title}{{Astropy: A community Python package for astronomy},} \aap, 558, A33, \dodoi{10.1051/0004-6361/201322068}

\bibitem[{ {Astropy Collaboration} {et~al.}(2018){Astropy Collaboration}, {Price-Whelan}, {Sip{\H{o}}cz}, {G{\"u}nther}, {Lim}, {Crawford}, {Conseil}, {Shupe}, {Craig}, {Dencheva}, {Ginsburg}, {VanderPlas}, {Bradley}, {P{\'e}rez-Su{\'a}rez}, {de Val-Borro}, {Aldcroft}, {Cruz}, {Robitaille}, {Tollerud}, {Ardelean}, {Babej}, {Bach}, {Bachetti}, {Bakanov}, {Bamford}, {Barentsen}, {Barmby}, {Baumbach}, {Berry}, {Biscani}, {Boquien}, {Bostroem}, {Bouma}, {Brammer}, {Bray}, {Breytenbach}, {Buddelmeijer}, {Burke}, {Calderone}, {Cano Rodr{\'\i}guez}, {Cara}, {Cardoso}, {Cheedella}, {Copin}, {Corrales}, {Crichton}, {D'Avella}, {Deil}, {Depagne}, {Dietrich}, {Donath}, {Droettboom}, {Earl}, {Erben}, {Fabbro}, {Ferreira}, {Finethy}, {Fox}, {Garrison}, {Gibbons}, {Goldstein}, {Gommers}, {Greco}, {Greenfield}, {Groener}, {Grollier}, {Hagen}, {Hirst}, {Homeier}, {Horton}, {Hosseinzadeh}, {Hu}, {Hunkeler}, {Ivezi{\'c}}, {Jain}, {Jenness}, {Kanarek}, {Kendrew}, {Kern}, {Kerzendorf}, {Khvalko}, {King}, {Kirkby}, {Kulkarni},
  {Kumar}, {Lee}, {Lenz}, {Littlefair}, {Ma}, {Macleod}, {Mastropietro}, {McCully}, {Montagnac}, {Morris}, {Mueller}, {Mumford}, {Muna}, {Murphy}, {Nelson}, {Nguyen}, {Ninan}, {N{\"o}the}, {Ogaz}, {Oh}, {Parejko}, {Parley}, {Pascual}, {Patil}, {Patil}, {Plunkett}, {Prochaska}, {Rastogi}, {Reddy Janga}, {Sabater}, {Sakurikar}, {Seifert}, {Sherbert}, {Sherwood-Taylor}, {Shih}, {Sick}, {Silbiger}, {Singanamalla}, {Singer}, {Sladen}, {Sooley}, {Sornarajah}, {Streicher}, {Teuben}, {Thomas}, {Tremblay}, {Turner}, {Terr{\'o}n}, {van Kerkwijk}, {de la Vega}, {Watkins}, {Weaver}, {Whitmore}, {Woillez}, {Zabalza}, \& {Astropy Contributors}}]{astropy+18}
{Astropy Collaboration}, {Price-Whelan}, A.~M., {Sip{\H{o}}cz}, B.~M., {et~al.} 2018, \bibinfo{title}{{The Astropy Project: Building an Open-science Project and Status of the v2.0 Core Package},} \aj, 156, 123, \dodoi{10.3847/1538-3881/aabc4f}

\bibitem[{ {Astropy Collaboration} {et~al.}(2022){Astropy Collaboration}, {Price-Whelan}, {Lim}, {Earl}, {Starkman}, {Bradley}, {Shupe}, {Patil}, {Corrales}, {Brasseur}, {N{\"o}the}, {Donath}, {Tollerud}, {Morris}, {Ginsburg}, {Vaher}, {Weaver}, {Tocknell}, {Jamieson}, {van Kerkwijk}, {Robitaille}, {Merry}, {Bachetti}, {G{\"u}nther}, {Aldcroft}, {Alvarado-Montes}, {Archibald}, {B{\'o}di}, {Bapat}, {Barentsen}, {Baz{\'a}n}, {Biswas}, {Boquien}, {Burke}, {Cara}, {Cara}, {Conroy}, {Conseil}, {Craig}, {Cross}, {Cruz}, {D'Eugenio}, {Dencheva}, {Devillepoix}, {Dietrich}, {Eigenbrot}, {Erben}, {Ferreira}, {Foreman-Mackey}, {Fox}, {Freij}, {Garg}, {Geda}, {Glattly}, {Gondhalekar}, {Gordon}, {Grant}, {Greenfield}, {Groener}, {Guest}, {Gurovich}, {Handberg}, {Hart}, {Hatfield-Dodds}, {Homeier}, {Hosseinzadeh}, {Jenness}, {Jones}, {Joseph}, {Kalmbach}, {Karamehmetoglu}, {Ka{\l}uszy{\'n}ski}, {Kelley}, {Kern}, {Kerzendorf}, {Koch}, {Kulumani}, {Lee}, {Ly}, {Ma}, {MacBride}, {Maljaars}, {Muna}, {Murphy}, {Norman},
  {O'Steen}, {Oman}, {Pacifici}, {Pascual}, {Pascual-Granado}, {Patil}, {Perren}, {Pickering}, {Rastogi}, {Roulston}, {Ryan}, {Rykoff}, {Sabater}, {Sakurikar}, {Salgado}, {Sanghi}, {Saunders}, {Savchenko}, {Schwardt}, {Seifert-Eckert}, {Shih}, {Jain}, {Shukla}, {Sick}, {Simpson}, {Singanamalla}, {Singer}, {Singhal}, {Sinha}, {Sip{\H{o}}cz}, {Spitler}, {Stansby}, {Streicher}, {{\v{S}}umak}, {Swinbank}, {Taranu}, {Tewary}, {Tremblay}, {de Val-Borro}, {Van Kooten}, {Vasovi{\'c}}, {Verma}, {de Miranda Cardoso}, {Williams}, {Wilson}, {Winkel}, {Wood-Vasey}, {Xue}, {Yoachim}, {Zhang}, {Zonca}, \& {Astropy Project Contributors}}]{astropy+22}
{Astropy Collaboration}, {Price-Whelan}, A.~M., {Lim}, P.~L., {et~al.} 2022, \bibinfo{title}{{The Astropy Project: Sustaining and Growing a Community-oriented Open-source Project and the Latest Major Release (v5.0) of the Core Package},} \apj, 935, 167, \dodoi{10.3847/1538-4357/ac7c74}

\bibitem[{M. {Bailes}(2022){Bailes}}]{Bailes22}
{Bailes}, M. 2022, \bibinfo{title}{{The discovery and scientific potential of fast radio bursts},} Science, 378, abj3043, \dodoi{10.1126/science.abj3043}

\bibitem[{K.~W. {Bannister} {et~al.}(2019){Bannister}, {Deller}, {Phillips}, {Macquart}, {Prochaska}, {Tejos}, {Ryder}, {Sadler}, {Shannon}, {Simha}, {Day}, {McQuinn}, {North-Hickey}, {Bhandari}, {Arcus}, {Bennert}, {Burchett}, {Bouwhuis}, {Dodson}, {Ekers}, {Farah}, {Flynn}, {James}, {Kerr}, {Lenc}, {Mahony}, {O'Meara}, {Os{\l}owski}, {Qiu}, {Treu}, {U}, {Bateman}, {Bock}, {Bolton}, {Brown}, {Bunton}, {Chippendale}, {Cooray}, {Cornwell}, {Gupta}, {Hayman}, {Kesteven}, {Koribalski}, {MacLeod}, {McClure-Griffiths}, {Neuhold}, {Norris}, {Pilawa}, {Qiao}, {Reynolds}, {Roxby}, {Shimwell}, {Voronkov}, \& {Wilson}}]{Bannister+19}
{Bannister}, K.~W., {Deller}, A.~T., {Phillips}, C., {et~al.} 2019, \bibinfo{title}{{A single fast radio burst localized to a massive galaxy at cosmological distance},} Science, 365, 565, \dodoi{10.1126/science.aaw5903}

\bibitem[{P. {Behroozi} {et~al.}(2019){Behroozi}, {Wechsler}, {Hearin}, \& {Conroy}}]{Behroozi+19}
{Behroozi}, P., {Wechsler}, R.~H., {Hearin}, A.~P., \& {Conroy}, C. 2019, \bibinfo{title}{{UNIVERSEMACHINE: The correlation between galaxy growth and dark matter halo assembly from z = 0-10},} \mnras, 488, 3143, \dodoi{10.1093/mnras/stz1182}

\bibitem[{A.~M. {Beloborodov}(2017){Beloborodov}}]{Beloborodov17}
{Beloborodov}, A.~M. 2017, \bibinfo{title}{{A Flaring Magnetar in FRB 121102?},} \apjl, 843, L26, \dodoi{10.3847/2041-8213/aa78f3}

\bibitem[{A.~M. {Beloborodov}(2020){Beloborodov}}]{Beloborodov20}
{Beloborodov}, A.~M. 2020, \bibinfo{title}{{Blast Waves from Magnetar Flares and Fast Radio Bursts},} \apj, 896, 142, \dodoi{10.3847/1538-4357/ab83eb}

\bibitem[{S. {Bhandari} {et~al.}(2022){Bhandari}, {Heintz}, {Aggarwal}, {Marnoch}, {Day}, {Sydnor}, {Burke-Spolaor}, {Law}, {Xavier Prochaska}, {Tejos}, {Bannister}, {Butler}, {Deller}, {Ekers}, {Flynn}, {Fong}, {James}, {Lazio}, {Luo}, {Mahony}, {Ryder}, {Sadler}, {Shannon}, {Han}, {Lee}, \& {Zhang}}]{Bhandari+22}
{Bhandari}, S., {Heintz}, K.~E., {Aggarwal}, K., {et~al.} 2022, \bibinfo{title}{{Characterizing the Fast Radio Burst Host Galaxy Population and its Connection to Transients in the Local and Extragalactic Universe},} \aj, 163, 69, \dodoi{10.3847/1538-3881/ac3aec}

\bibitem[{M. {Bhardwaj} {et~al.}(2024){Bhardwaj}, {Michilli}, {Kirichenko}, {Modilim}, {Shin}, {Kaspi}, {Andersen}, {Cassanelli}, {Brar}, {Chatterjee}, {Cook}, {Dong}, {Fonseca}, {Gaensler}, {Ibik}, {Kaczmarek}, {Lanman}, {Leung}, {Masui}, {Pandhi}, {Pearlman}, {Petroff}, {Pleunis}, {Prochaska}, {Rafiei-Ravandi}, {Sand}, {Scholz}, \& {Smith}}]{Bhardwaj+24}
{Bhardwaj}, M., {Michilli}, D., {Kirichenko}, A.~Y., {et~al.} 2024, \bibinfo{title}{{Host Galaxies for Four Nearby CHIME/FRB Sources and the Local Universe FRB Host Galaxy Population},} \apjl, 971, L51, \dodoi{10.3847/2041-8213/ad64d1}

\bibitem[{C.~D. {Bochenek} {et~al.}(2020){Bochenek}, {Ravi}, {Belov}, {Hallinan}, {Kocz}, {Kulkarni}, \& {McKenna}}]{Bochenek+20}
{Bochenek}, C.~D., {Ravi}, V., {Belov}, K.~V., {et~al.} 2020, \bibinfo{title}{{A fast radio burst associated with a Galactic magnetar},} \nat, 587, 59, \dodoi{10.1038/s41586-020-2872-x}

\bibitem[{C.~D. {Bochenek} {et~al.}(2021){Bochenek}, {Ravi}, \& {Dong}}]{Bochenek+21}
{Bochenek}, C.~D., {Ravi}, V., \& {Dong}, D. 2021, \bibinfo{title}{{Localized Fast Radio Bursts Are Consistent with Magnetar Progenitors Formed in Core-collapse Supernovae},} \apjl, 907, L31, \dodoi{10.3847/2041-8213/abd634}

\bibitem[{J.~H. {Chen} {et~al.}(2024){Chen}, {Jia}, {Dong}, \& {Wang}}]{Chen+24}
{Chen}, J.~H., {Jia}, X.~D., {Dong}, X.~F., \& {Wang}, F.~Y. 2024, \bibinfo{title}{{The Formation Rate and Luminosity Function of Fast Radio Bursts},} \apjl, 973, L54, \dodoi{10.3847/2041-8213/ad7b39}

\bibitem[{ {CHIME/FRB Collaboration} {et~al.}(2020){CHIME/FRB Collaboration}, {Andersen}, {Bandura}, {Bhardwaj}, {Bij}, {Boyce}, {Boyle}, {Brar}, {Cassanelli}, {Chawla}, {Chen}, {Cliche}, {Cook}, {Cubranic}, {Curtin}, {Denman}, {Dobbs}, {Dong}, {Fandino}, {Fonseca}, {Gaensler}, {Giri}, {Good}, {Halpern}, {Hill}, {Hinshaw}, {H{\"o}fer}, {Josephy}, {Kania}, {Kaspi}, {Landecker}, {Leung}, {Li}, {Lin}, {Masui}, {McKinven}, {Mena-Parra}, {Merryfield}, {Meyers}, {Michilli}, {Milutinovic}, {Mirhosseini}, {M{\"u}nchmeyer}, {Naidu}, {Newburgh}, {Ng}, {Patel}, {Pen}, {Pinsonneault-Marotte}, {Pleunis}, {Quine}, {Rafiei-Ravandi}, {Rahman}, {Ransom}, {Renard}, {Sanghavi}, {Scholz}, {Shaw}, {Shin}, {Siegel}, {Singh}, {Smegal}, {Smith}, {Stairs}, {Tan}, {Tendulkar}, {Tretyakov}, {Vanderlinde}, {Wang}, {Wulf}, \& {Zwaniga}}]{Andersen+20}
{CHIME/FRB Collaboration}, {Andersen}, B.~C., {Bandura}, K.~M., {et~al.} 2020, \bibinfo{title}{{A bright millisecond-duration radio burst from a Galactic magnetar},} \nat, 587, 54, \dodoi{10.1038/s41586-020-2863-y}

\bibitem[{L. {Connor} {et~al.}(2023){Connor}, {Ravi}, {Catha}, {Chen}, {Faber}, {Lamb}, {Hallinan}, {Harnach}, {Hellbourg}, {Hobbs}, {Hodge}, {Hodges}, {Law}, {Rasmussen}, {Sayers}, {Sharma}, {Sherman}, {Shi}, {Simard}, {Somalwar}, {Squillace}, {Weinreb}, {Woody}, {Yadlapalli}, \& {Deep Synoptic Array Team}}]{Connor+23}
{Connor}, L., {Ravi}, V., {Catha}, M., {et~al.} 2023, \bibinfo{title}{{Deep Synoptic Array Science: Two Fast Radio Burst Sources in Massive Galaxy Clusters},} \apjl, 949, L26, \dodoi{10.3847/2041-8213/acd3ea}

\bibitem[{J.~M. {Cordes} \& S. {Chatterjee}(2019){Cordes} \& {Chatterjee}}]{Cordes&Chatterjee19}
{Cordes}, J.~M., \& {Chatterjee}, S. 2019, \bibinfo{title}{{Fast Radio Bursts: An Extragalactic Enigma},} \araa, 57, 417, \dodoi{10.1146/annurev-astro-091918-104501}

\bibitem[{G.~M. {Eadie} {et~al.}(2022){Eadie}, {Harris}, \& {Springford}}]{Eadie+22}
{Eadie}, G.~M., {Harris}, W.~E., \& {Springford}, A. 2022, \bibinfo{title}{{Clearing the Hurdle: The Mass of Globular Cluster Systems as a Function of Host Galaxy Mass},} \apj, 926, 162, \dodoi{10.3847/1538-4357/ac33b0}

\bibitem[{T. {Eftekhari} {et~al.}(2025){Eftekhari}, {Dong}, {Fong}, {Shah}, {Simha}, {Andersen}, {Andrew}, {Bhardwaj}, {Cassanelli}, {Chatterjee}, {Coulter}, {Fonseca}, {Gaensler}, {Gordon}, {Hessels}, {Ibik}, {Joseph}, {Kahinga}, {Kaspi}, {Kharel}, {Kilpatrick}, {Lanman}, {Lazda}, {Leung}, {Liu}, {Mas-Ribas}, {Masui}, {Mckinven}, {Mena-Parra}, {Miller}, {Nimmo}, {Pandhi}, {Patil}, {Pearlman}, {Pleunis}, {Prochaska}, {Rafiei-Ravandi}, {Sammons}, {Scholz}, {Shin}, {Smith}, \& {Stairs}}]{Eftekhari+25}
{Eftekhari}, T., {Dong}, Y., {Fong}, W., {et~al.} 2025, \bibinfo{title}{{The Massive and Quiescent Elliptical Host Galaxy of the Repeating Fast Radio Burst FRB 20240209A},} \apjl, 979, L22, \dodoi{10.3847/2041-8213/ad9de2}

\bibitem[{G. {Fasano} \& A. {Franceschini}(1987){Fasano} \& {Franceschini}}]{Fasano&Franceschini87}
{Fasano}, G., \& {Franceschini}, A. 1987, \bibinfo{title}{{A multidimensional version of the Kolmogorov-Smirnov test},} \mnras, 225, 155, \dodoi{10.1093/mnras/225.1.155}

\bibitem[{W. {Fong} {et~al.}(2013){Fong}, {Berger}, {Chornock}, {Margutti}, {Levan}, {Tanvir}, {Tunnicliffe}, {Czekala}, {Fox}, {Perley}, {Cenko}, {Zauderer}, {Laskar}, {Persson}, {Monson}, {Kelson}, {Birk}, {Murphy}, {Servillat}, \& {Anglada}}]{Fong+13}
{Fong}, W., {Berger}, E., {Chornock}, R., {et~al.} 2013, \bibinfo{title}{{Demographics of the Galaxies Hosting Short-duration Gamma-Ray Bursts},} \apj, 769, 56, \dodoi{10.1088/0004-637X/769/1/56}

\bibitem[{D. {Foreman-Mackey} {et~al.}(2013){Foreman-Mackey}, {Hogg}, {Lang}, \& {Goodman}}]{Foreman-Mackey+13}
{Foreman-Mackey}, D., {Hogg}, D.~W., {Lang}, D., \& {Goodman}, J. 2013, \bibinfo{title}{{emcee: The MCMC Hammer},} \pasp, 125, 306, \dodoi{10.1086/670067}

\bibitem[{A.~S. {Fruchter} {et~al.}(2006){Fruchter}, {Levan}, {Strolger}, {Vreeswijk}, {Thorsett}, {Bersier}, {Burud}, {Castro Cer{\'o}n}, {Castro-Tirado}, {Conselice}, {Dahlen}, {Ferguson}, {Fynbo}, {Garnavich}, {Gibbons}, {Gorosabel}, {Gull}, {Hjorth}, {Holland}, {Kouveliotou}, {Levay}, {Livio}, {Metzger}, {Nugent}, {Petro}, {Pian}, {Rhoads}, {Riess}, {Sahu}, {Smette}, {Tanvir}, {Wijers}, \& {Woosley}}]{Fruchter+06}
{Fruchter}, A.~S., {Levan}, A.~J., {Strolger}, L., {et~al.} 2006, \bibinfo{title}{{Long {\ensuremath{\gamma}}-ray bursts and core-collapse supernovae have different environments},} \nat, 441, 463, \dodoi{10.1038/nature04787}

\bibitem[{A.~C. {Gordon} {et~al.}(2023){Gordon}, {Fong}, {Kilpatrick}, {Eftekhari}, {Leja}, {Prochaska}, {Nugent}, {Bhandari}, {Blanchard}, {Caleb}, {Day}, {Deller}, {Dong}, {Glowacki}, {Gourdji}, {Mannings}, {Mahoney}, {Marnoch}, {Miller}, {Paterson}, {Rastinejad}, {Ryder}, {Sadler}, {Scott}, {Sears}, {Shannon}, {Simha}, {Stappers}, \& {Tejos}}]{Gordon+23}
{Gordon}, A.~C., {Fong}, W.-f., {Kilpatrick}, C.~D., {et~al.} 2023, \bibinfo{title}{{The Demographics, Stellar Populations, and Star Formation Histories of Fast Radio Burst Host Galaxies: Implications for the Progenitors},} \apj, 954, 80, \dodoi{10.3847/1538-4357/ace5aa}

\bibitem[{A.~C. {Gordon} {et~al.}(2025){Gordon}, {Fong}, {Deller}, {Marnoch}, {Lim}, {Peng}, {Bannister}, {Bera}, {Bhat}, {Dial}, {Dong}, {Eftekhari}, {Glowacki}, {Gourdji}, {Gupta}, {Jahns-Schindler}, {Jaini}, {Kilpatrick}, {Liu}, {Prochaska}, {Ryder}, {Shannon}, {Simha}, {Tejos}, {Wang}, \& {Wang}}]{Gordon+25}
{Gordon}, A.~C., {Fong}, W.-f., {Deller}, A.~T., {et~al.} 2025, \bibinfo{title}{{Mapping the Spatial Distribution of Fast Radio Bursts within their Host Galaxies},} arXiv e-prints, arXiv:2506.06453, \dodoi{10.48550/arXiv.2506.06453}

\bibitem[{O. {Gupta} {et~al.}(2025){Gupta}, {Beniamini}, {Kumar}, \& {Finkelstein}}]{Gupta+25}
{Gupta}, O., {Beniamini}, P., {Kumar}, P., \& {Finkelstein}, S.~L. 2025, \bibinfo{title}{{The Cosmic Evolution of Fast Radio Bursts Inferred from the CHIME/FRB Baseband Catalog 1},} \apj, 986, 100, \dodoi{10.3847/1538-4357/add14c}

\bibitem[{C.~R. Harris {et~al.}(2020)Harris, Millman, van~der Walt, Gommers, Virtanen, Cournapeau, Wieser, Taylor, Berg, Smith, Kern, Picus, Hoyer, van Kerkwijk, Brett, Haldane, del R{\'{i}}o, Wiebe, Peterson, G{\'{e}}rard-Marchant, Sheppard, Reddy, Weckesser, Abbasi, Gohlke, \& Oliphant}]{numpy}
Harris, C.~R., Millman, K.~J., van~der Walt, S.~J., {et~al.} 2020, \bibinfo{title}{Array programming with {NumPy},} Nature, 585, 357, \dodoi{10.1038/s41586-020-2649-2}

\bibitem[{W.~E. {Harris} {et~al.}(2017){Harris}, {Blakeslee}, \& {Harris}}]{Harris+17}
{Harris}, W.~E., {Blakeslee}, J.~P., \& {Harris}, G. L.~H. 2017, \bibinfo{title}{{Galactic Dark Matter Halos and Globular Cluster Populations. III. Extension to Extreme Environments},} \apj, 836, 67, \dodoi{10.3847/1538-4357/836/1/67}

\bibitem[{W.~E. {Harris} {et~al.}(2013){Harris}, {Harris}, \& {Alessi}}]{Harris+13}
{Harris}, W.~E., {Harris}, G. L.~H., \& {Alessi}, M. 2013, \bibinfo{title}{{A Catalog of Globular Cluster Systems: What Determines the Size of a Galaxy's Globular Cluster Population?},} \apj, 772, 82, \dodoi{10.1088/0004-637X/772/2/82}

\bibitem[{T. {Hashimoto} {et~al.}(2022){Hashimoto}, {Goto}, {Chen}, {Ho}, {Hsiao}, {Wong}, {On}, {Kim}, {Kilerci-Eser}, {Huang}, {Santos}, \& {Yamasaki}}]{Hashimoto+22}
{Hashimoto}, T., {Goto}, T., {Chen}, B.~H., {et~al.} 2022, \bibinfo{title}{{Energy functions of fast radio bursts derived from the first CHIME/FRB catalogue},} \mnras, 511, 1961, \dodoi{10.1093/mnras/stac065}

\bibitem[{K.~E. {Heintz} {et~al.}(2020){Heintz}, {Prochaska}, {Simha}, {Platts}, {Fong}, {Tejos}, {Ryder}, {Aggerwal}, {Bhandari}, {Day}, {Deller}, {Kilpatrick}, {Law}, {Macquart}, {Mannings}, {Marnoch}, {Sadler}, \& {Shannon}}]{Heintz+20}
{Heintz}, K.~E., {Prochaska}, J.~X., {Simha}, S., {et~al.} 2020, \bibinfo{title}{{Host Galaxy Properties and Offset Distributions of Fast Radio Bursts: Implications for Their Progenitors},} \apj, 903, 152, \dodoi{10.3847/1538-4357/abb6fb}

\bibitem[{A. Horowicz(2025)Horowicz}]{host_galaxies_stats}
Horowicz, A. 2025, \bibinfo{title}{Host Galaxies Stats - a python framework for investigating host-galaxxies population of astrophysical transients,}, 1.0 Zenodo, \dodoi{10.5281/zenodo.17415056}

\bibitem[{M.~J. {Hudson} {et~al.}(2014){Hudson}, {Harris}, \& {Harris}}]{Hudson+14}
{Hudson}, M.~J., {Harris}, G.~L., \& {Harris}, W.~E. 2014, \bibinfo{title}{{Dark Matter Halos in Galaxies and Globular Cluster Populations},} \apjl, 787, L5, \dodoi{10.1088/2041-8205/787/1/L5}

\bibitem[{J.~D. Hunter(2007)Hunter}]{matplotlib}
Hunter, J.~D. 2007, \bibinfo{title}{Matplotlib: A 2D graphics environment,} Computing in Science \& Engineering, 9, 90, \dodoi{10.1109/MCSE.2007.55}

\bibitem[{C.~W. {James} {et~al.}(2022){James}, {Prochaska}, {Macquart}, {North-Hickey}, {Bannister}, \& {Dunning}}]{James+22}
{James}, C.~W., {Prochaska}, J.~X., {Macquart}, J.-P., {et~al.} 2022, \bibinfo{title}{{The fast radio burst population evolves, consistent with the star formation rate},} \mnras, 510, L18, \dodoi{10.1093/mnrasl/slab117}

\bibitem[{B.~D. {Johnson} {et~al.}(2021){Johnson}, {Leja}, {Conroy}, \& {Speagle}}]{Johnson+21}
{Johnson}, B.~D., {Leja}, J., {Conroy}, C., \& {Speagle}, J.~S. 2021, \bibinfo{title}{{Stellar Population Inference with Prospector},} \apjs, 254, 22, \dodoi{10.3847/1538-4365/abef67}

\bibitem[{K. {Kashiyama} \& K. {Murase}(2017){Kashiyama} \& {Murase}}]{Kashiyama&Murase17}
{Kashiyama}, K., \& {Murase}, K. 2017, \bibinfo{title}{{Testing the Young Neutron Star Scenario with Persistent Radio Emission Associated with FRB 121102},} \apjl, 839, L3, \dodoi{10.3847/2041-8213/aa68e1}

\bibitem[{J.~I. {Katz}(2016){Katz}}]{Katz16}
{Katz}, J.~I. 2016, \bibinfo{title}{{How Soft Gamma Repeaters Might Make Fast Radio Bursts},} \apj, 826, 226, \dodoi{10.3847/0004-637X/826/2/226}

\bibitem[{F. {Kirsten} {et~al.}(2022){Kirsten}, {Marcote}, {Nimmo}, {Hessels}, {Bhardwaj}, {Tendulkar}, {Keimpema}, {Yang}, {Snelders}, {Scholz}, {Pearlman}, {Law}, {Peters}, {Giroletti}, {Paragi}, {Bassa}, {Hewitt}, {Bach}, {Bezrukovs}, {Burgay}, {Buttaccio}, {Conway}, {Corongiu}, {Feiler}, {Forss{\'e}n}, {Gawro{\'n}ski}, {Karuppusamy}, {Kharinov}, {Lindqvist}, {Maccaferri}, {Melnikov}, {Ould-Boukattine}, {Possenti}, {Surcis}, {Wang}, {Yuan}, {Aggarwal}, {Anna-Thomas}, {Bower}, {Blaauw}, {Burke-Spolaor}, {Cassanelli}, {Clarke}, {Fonseca}, {Gaensler}, {Gopinath}, {Kaspi}, {Kassim}, {Lazio}, {Leung}, {Li}, {Lin}, {Masui}, {Mckinven}, {Michilli}, {Mikhailov}, {Ng}, {Orbidans}, {Pen}, {Petroff}, {Rahman}, {Ransom}, {Shin}, {Smith}, {Stairs}, \& {Vlemmings}}]{Kirsten+22}
{Kirsten}, F., {Marcote}, B., {Nimmo}, K., {et~al.} 2022, \bibinfo{title}{{A repeating fast radio burst source in a globular cluster},} \nat, 602, 585, \dodoi{10.1038/s41586-021-04354-w}

\bibitem[{K. {Kremer} {et~al.}(2021){Kremer}, {Piro}, \& {Li}}]{Kremer+21}
{Kremer}, K., {Piro}, A.~L., \& {Li}, D. 2021, \bibinfo{title}{{Dynamical Formation Channels for Fast Radio Bursts in Globular Clusters},} \apjl, 917, L11, \dodoi{10.3847/2041-8213/ac13a0}

\bibitem[{S.~R. {Kulkarni} {et~al.}(2014){Kulkarni}, {Ofek}, {Neill}, {Zheng}, \& {Juric}}]{Kulkarni+14}
{Kulkarni}, S.~R., {Ofek}, E.~O., {Neill}, J.~D., {Zheng}, Z., \& {Juric}, M. 2014, \bibinfo{title}{{Giant Sparks at Cosmological Distances?},} \apj, 797, 70, \dodoi{10.1088/0004-637X/797/1/70}

\bibitem[{P. {Kumar} \& {\v{Z}}. {Bo{\v{s}}njak}(2020){Kumar} \& {Bo{\v{s}}njak}}]{Kumar&Bosnjak20}
{Kumar}, P., \& {Bo{\v{s}}njak}, {\v{Z}}. 2020, \bibinfo{title}{{FRB coherent emission from decay of Alfv{\'e}n waves},} \mnras, 494, 2385, \dodoi{10.1093/mnras/staa774}

\bibitem[{P. {Kumar} {et~al.}(2017){Kumar}, {Lu}, \& {Bhattacharya}}]{Kumar+17}
{Kumar}, P., {Lu}, W., \& {Bhattacharya}, M. 2017, \bibinfo{title}{{Fast radio burst source properties and curvature radiation model},} \mnras, 468, 2726, \dodoi{10.1093/mnras/stx665}

\bibitem[{C. {Laigle} {et~al.}(2016){Laigle}, {McCracken}, {Ilbert}, {Hsieh}, {Davidzon}, {Capak}, {Hasinger}, {Silverman}, {Pichon}, {Coupon}, {Aussel}, {Le Borgne}, {Caputi}, {Cassata}, {Chang}, {Civano}, {Dunlop}, {Fynbo}, {Kartaltepe}, {Koekemoer}, {Le F{\`e}vre}, {Le Floc'h}, {Leauthaud}, {Lilly}, {Lin}, {Marchesi}, {Milvang-Jensen}, {Salvato}, {Sanders}, {Scoville}, {Smolcic}, {Stockmann}, {Taniguchi}, {Tasca}, {Toft}, {Vaccari}, \& {Zabl}}]{Laigle+16}
{Laigle}, C., {McCracken}, H.~J., {Ilbert}, O., {et~al.} 2016, \bibinfo{title}{{The COSMOS2015 Catalog: Exploring the 1 < z < 6 Universe with Half a Million Galaxies},} \apjs, 224, 24, \dodoi{10.3847/0067-0049/224/2/24}

\bibitem[{A.~E. {Lanman} {et~al.}(2024){Lanman}, {Andrew}, {Lazda}, {Shah}, {Amiri}, {Balasubramanian}, {Bandura}, {Boyle}, {Brar}, {Carlson}, {Cliche}, {Gusinskaia}, {Hendricksen}, {Kaczmarek}, {Landecker}, {Leung}, {Mckinven}, {Mena-Parra}, {Milutinovic}, {Nimmo}, {Pearlman}, {Renard}, {Rahman}, {Shaw}, {Siegel}, {Smegal}, {Cassanelli}, {Chatterjee}, {Curtin}, {Dobbs}, {Dong}, {Halpern}, {Hopkins}, {Kaspi}, {Khairy}, {Masui}, {Meyers}, {Michilli}, {Petroff}, {Pinsonneault-Marotte}, {Pleunis}, {Rafiei-Ravandi}, {Shin}, {Smith}, {Vanderlinde}, \& {Zegmott}}]{Lanman+24}
{Lanman}, A.~E., {Andrew}, S., {Lazda}, M., {et~al.} 2024, \bibinfo{title}{{CHIME/FRB Outriggers: KKO Station System and Commissioning Results},} \aj, 168, 87, \dodoi{10.3847/1538-3881/ad5838}

\bibitem[{C.~J. {Law} {et~al.}(2024){Law}, {Sharma}, {Ravi}, {Chen}, {Catha}, {Connor}, {Faber}, {Hallinan}, {Harnach}, {Hellbourg}, {Hobbs}, {Hodge}, {Hodges}, {Lamb}, {Rasmussen}, {Sherman}, {Shi}, {Simard}, {Squillace}, {Weinreb}, {Woody}, \& {Yurk}}]{Law+24}
{Law}, C.~J., {Sharma}, K., {Ravi}, V., {et~al.} 2024, \bibinfo{title}{{Deep Synoptic Array Science: First FRB and Host Galaxy Catalog},} \apj, 967, 29, \dodoi{10.3847/1538-4357/ad3736}

\bibitem[{J. {Leja} {et~al.}(2017){Leja}, {Johnson}, {Conroy}, {van Dokkum}, \& {Byler}}]{Leja+17}
{Leja}, J., {Johnson}, B.~D., {Conroy}, C., {van Dokkum}, P.~G., \& {Byler}, N. 2017, \bibinfo{title}{{Deriving Physical Properties from Broadband Photometry with Prospector: Description of the Model and a Demonstration of its Accuracy Using 129 Galaxies in the Local Universe},} \apj, 837, 170, \dodoi{10.3847/1538-4357/aa5ffe}

\bibitem[{J. {Leja} {et~al.}(2020){Leja}, {Speagle}, {Johnson}, {Conroy}, {van Dokkum}, \& {Franx}}]{Leja+20}
{Leja}, J., {Speagle}, J.~S., {Johnson}, B.~D., {et~al.} 2020, \bibinfo{title}{{A New Census of the 0.2 < z < 3.0 Universe. I. The Stellar Mass Function},} \apj, 893, 111, \dodoi{10.3847/1538-4357/ab7e27}

\bibitem[{J. {Leja} {et~al.}(2019){Leja}, {Johnson}, {Conroy}, {van Dokkum}, {Speagle}, {Brammer}, {Momcheva}, {Skelton}, {Whitaker}, {Franx}, \& {Nelson}}]{Leja+19}
{Leja}, J., {Johnson}, B.~D., {Conroy}, C., {et~al.} 2019, \bibinfo{title}{{An Older, More Quiescent Universe from Panchromatic SED Fitting of the 3D-HST Survey},} \apj, 877, 140, \dodoi{10.3847/1538-4357/ab1d5a}

\bibitem[{J. {Leja} {et~al.}(2022){Leja}, {Speagle}, {Ting}, {Johnson}, {Conroy}, {Whitaker}, {Nelson}, {van Dokkum}, \& {Franx}}]{Leja+22}
{Leja}, J., {Speagle}, J.~S., {Ting}, Y.-S., {et~al.} 2022, \bibinfo{title}{{A New Census of the 0.2 < z < 3.0 Universe. II. The Star-forming Sequence},} \apj, 936, 165, \dodoi{10.3847/1538-4357/ac887d}

\bibitem[{D.~R. {Lorimer} {et~al.}(2007){Lorimer}, {Bailes}, {McLaughlin}, {Narkevic}, \& {Crawford}}]{Lorimer+07}
{Lorimer}, D.~R., {Bailes}, M., {McLaughlin}, M.~A., {Narkevic}, D.~J., \& {Crawford}, F. 2007, \bibinfo{title}{{A Bright Millisecond Radio Burst of Extragalactic Origin},} Science, 318, 777, \dodoi{10.1126/science.1147532}

\bibitem[{N. {Loudas} {et~al.}(2025){Loudas}, {Li}, {Strauss}, \& {Leja}}]{Loudas+25}
{Loudas}, N., {Li}, D., {Strauss}, M.~A., \& {Leja}, J. 2025, \bibinfo{title}{{Unveiling the origin of fast radio bursts by modeling the stellar mass and star formation distributions of their host galaxies},} arXiv e-prints, arXiv:2502.15566, \dodoi{10.48550/arXiv.2502.15566}

\bibitem[{W. {Lu} {et~al.}(2022){Lu}, {Beniamini}, \& {Kumar}}]{Lu+22}
{Lu}, W., {Beniamini}, P., \& {Kumar}, P. 2022, \bibinfo{title}{{Implications of a rapidly varying FRB in a globular cluster of M81},} \mnras, 510, 1867, \dodoi{10.1093/mnras/stab3500}

\bibitem[{W. {Lu} \& P. {Kumar}(2018){Lu} \& {Kumar}}]{Lu&Kumar18}
{Lu}, W., \& {Kumar}, P. 2018, \bibinfo{title}{{On the radiation mechanism of repeating fast radio bursts},} \mnras, 477, 2470, \dodoi{10.1093/mnras/sty716}

\bibitem[{W. {Lu} {et~al.}(2020){Lu}, {Kumar}, \& {Zhang}}]{Lu+20}
{Lu}, W., {Kumar}, P., \& {Zhang}, B. 2020, \bibinfo{title}{{A unified picture of Galactic and cosmological fast radio bursts},} \mnras, 498, 1397, \dodoi{10.1093/mnras/staa2450}

\bibitem[{R. {Lunnan} {et~al.}(2014){Lunnan}, {Chornock}, {Berger}, {Laskar}, {Fong}, {Rest}, {Sanders}, {Challis}, {Drout}, {Foley}, {Huber}, {Kirshner}, {Leibler}, {Marion}, {McCrum}, {Milisavljevic}, {Narayan}, {Scolnic}, {Smartt}, {Smith}, {Soderberg}, {Tonry}, {Burgett}, {Chambers}, {Flewelling}, {Hodapp}, {Kaiser}, {Magnier}, {Price}, \& {Wainscoat}}]{Lunnan+14}
{Lunnan}, R., {Chornock}, R., {Berger}, E., {et~al.} 2014, \bibinfo{title}{{Hydrogen-poor Superluminous Supernovae and Long-duration Gamma-Ray Bursts Have Similar Host Galaxies},} \apj, 787, 138, \dodoi{10.1088/0004-637X/787/2/138}

\bibitem[{Y. {Lyubarsky}(2014){Lyubarsky}}]{Lyubarsky+14}
{Lyubarsky}, Y. 2014, \bibinfo{title}{{A model for fast extragalactic radio bursts.},} \mnras, 442, L9, \dodoi{10.1093/mnrasl/slu046}

\bibitem[{P. {Madau} \& M. {Dickinson}(2014){Madau} \& {Dickinson}}]{MadauDickinson14}
{Madau}, P., \& {Dickinson}, M. 2014, \bibinfo{title}{{Cosmic Star-Formation History},} \araa, 52, 415, \dodoi{10.1146/annurev-astro-081811-125615}

\bibitem[{A.~G. {Mannings} {et~al.}(2021){Mannings}, {Fong}, {Simha}, {Prochaska}, {Rafelski}, {Kilpatrick}, {Tejos}, {Heintz}, {Bannister}, {Bhandari}, {Day}, {Deller}, {Ryder}, {Shannon}, \& {Tendulkar}}]{Mannings+21}
{Mannings}, A.~G., {Fong}, W.-f., {Simha}, S., {et~al.} 2021, \bibinfo{title}{{A High-resolution View of Fast Radio Burst Host Environments},} \apj, 917, 75, \dodoi{10.3847/1538-4357/abff56}

\bibitem[{B. {Margalit} {et~al.}(2020){Margalit}, {Beniamini}, {Sridhar}, \& {Metzger}}]{Margalit+20}
{Margalit}, B., {Beniamini}, P., {Sridhar}, N., \& {Metzger}, B.~D. 2020, \bibinfo{title}{{Implications of a Fast Radio Burst from a Galactic Magnetar},} \apjl, 899, L27, \dodoi{10.3847/2041-8213/abac57}

\bibitem[{B. {Margalit} {et~al.}(2019){Margalit}, {Berger}, \& {Metzger}}]{Margalit+19}
{Margalit}, B., {Berger}, E., \& {Metzger}, B.~D. 2019, \bibinfo{title}{{Fast Radio Bursts from Magnetars Born in Binary Neutron Star Mergers and Accretion Induced Collapse},} \apj, 886, 110, \dodoi{10.3847/1538-4357/ab4c31}

\bibitem[{B. {Margalit} \& B.~D. {Metzger}(2018){Margalit} \& {Metzger}}]{Margalit+18}
{Margalit}, B., \& {Metzger}, B.~D. 2018, \bibinfo{title}{{A Concordance Picture of FRB 121102 as a Flaring Magnetar Embedded in a Magnetized Ion-Electron Wind Nebula},} \apjl, 868, L4, \dodoi{10.3847/2041-8213/aaedad}

\bibitem[{B.~D. {Metzger} {et~al.}(2017){Metzger}, {Berger}, \& {Margalit}}]{Metzger+17}
{Metzger}, B.~D., {Berger}, E., \& {Margalit}, B. 2017, \bibinfo{title}{{Millisecond Magnetar Birth Connects FRB 121102 to Superluminous Supernovae and Long-duration Gamma-Ray Bursts},} \apj, 841, 14, \dodoi{10.3847/1538-4357/aa633d}

\bibitem[{J.-f. {Mo} {et~al.}(2025){Mo}, {Zhu}, \& {Feng}}]{Mo+25}
{Mo}, J.-f., {Zhu}, W., \& {Feng}, L.-L. 2025, \bibinfo{title}{{The Dispersion Measure and Scattering of Fast Radio Bursts: Contributions from Multicomponents, and Clues for the Intrinsic Properties},} \apjs, 277, 43, \dodoi{10.3847/1538-4365/adb616}

\bibitem[{K. {Murase} {et~al.}(2016){Murase}, {Kashiyama}, \& {M{\'e}sz{\'a}ros}}]{Murase+16}
{Murase}, K., {Kashiyama}, K., \& {M{\'e}sz{\'a}ros}, P. 2016, \bibinfo{title}{{A burst in a wind bubble and the impact on baryonic ejecta: high-energy gamma-ray flashes and afterglows from fast radio bursts and pulsar-driven supernova remnants},} \mnras, 461, 1498, \dodoi{10.1093/mnras/stw1328}

\bibitem[{A.~E. {Nugent} {et~al.}(2025){Nugent}, {Villar}, {Gagliano}, {Jones}, {Horowicz}, {de Soto}, {Wang}, \& {Margalit}}]{Nugent+25}
{Nugent}, A.~E., {Villar}, V.~A., {Gagliano}, A., {et~al.} 2025, \bibinfo{title}{{Characterizing Supernovae Host Galaxies with FrankenBlast: A Scalable Tool for Transient Host Galaxy Association, Photometry, and Stellar Population Modeling},} arXiv e-prints, arXiv:2509.08874, \dodoi{10.48550/arXiv.2509.08874}

\bibitem[{A.~E. {Nugent} {et~al.}(2022){Nugent}, {Fong}, {Dong}, {Leja}, {Berger}, {Zevin}, {Chornock}, {Cobb}, {Kelley}, {Kilpatrick}, {Levan}, {Margutti}, {Paterson}, {Perley}, {Escorial}, {Smith}, \& {Tanvir}}]{Nugent+22}
{Nugent}, A.~E., {Fong}, W.-F., {Dong}, Y., {et~al.} 2022, \bibinfo{title}{{Short GRB Host Galaxies. II. A Legacy Sample of Redshifts, Stellar Population Properties, and Implications for Their Neutron Star Merger Origins},} \apj, 940, 57, \dodoi{10.3847/1538-4357/ac91d1}

\bibitem[{J.~A. {Peacock}(1983){Peacock}}]{Peacock83}
{Peacock}, J.~A. 1983, \bibinfo{title}{{Two-dimensional goodness-of-fit testing in astronomy.},} \mnras, 202, 615, \dodoi{10.1093/mnras/202.3.615}

\bibitem[{E. {Petroff} {et~al.}(2022){Petroff}, {Hessels}, \& {Lorimer}}]{Petroff+22}
{Petroff}, E., {Hessels}, J.~W.~T., \& {Lorimer}, D.~R. 2022, \bibinfo{title}{{Fast radio bursts at the dawn of the 2020s},} \aapr, 30, 2, \dodoi{10.1007/s00159-022-00139-w}

\bibitem[{D. {Pooley} {et~al.}(2003){Pooley}, {Lewin}, {Anderson}, {Baumgardt}, {Filippenko}, {Gaensler}, {Homer}, {Hut}, {Kaspi}, {Makino}, {Margon}, {McMillan}, {Portegies Zwart}, {van der Klis}, \& {Verbunt}}]{Pooley+03}
{Pooley}, D., {Lewin}, W. H.~G., {Anderson}, S.~F., {et~al.} 2003, \bibinfo{title}{{Dynamical Formation of Close Binary Systems in Globular Clusters},} \apjl, 591, L131, \dodoi{10.1086/377074}

\bibitem[{S.~B. {Popov} \& K.~A. {Postnov}(2013){Popov} \& {Postnov}}]{Popov+13}
{Popov}, S.~B., \& {Postnov}, K.~A. 2013, \bibinfo{title}{{Millisecond extragalactic radio bursts as magnetar flares},} arXiv e-prints, arXiv:1307.4924, \dodoi{10.48550/arXiv.1307.4924}

\bibitem[{A. {Rao} {et~al.}(2025){Rao}, {Ye}, \& {Fishbach}}]{Rao+25}
{Rao}, A., {Ye}, C.~S., \& {Fishbach}, M. 2025, \bibinfo{title}{{Predicting the Rate of Fast Radio Bursts in Globular Clusters from Binary Black Hole Observations},} \apjl, 979, L12, \dodoi{10.3847/2041-8213/ad9f2e}

\bibitem[{V. {Ravi} {et~al.}(2019){Ravi}, {Catha}, {D'Addario}, {Djorgovski}, {Hallinan}, {Hobbs}, {Kocz}, {Kulkarni}, {Shi}, {Vedantham}, {Weinreb}, \& {Woody}}]{Ravi+19}
{Ravi}, V., {Catha}, M., {D'Addario}, L., {et~al.} 2019, \bibinfo{title}{{A fast radio burst localized to a massive galaxy},} \nat, 572, 352, \dodoi{10.1038/s41586-019-1389-7}

\bibitem[{V. {Ravi} {et~al.}(2023){Ravi}, {Catha}, {Chen}, {Connor}, {Faber}, {Lamb}, {Hallinan}, {Harnach}, {Hellbourg}, {Hobbs}, {Hodge}, {Hodges}, {Law}, {Rasmussen}, {Sharma}, {Sherman}, {Shi}, {Simard}, {Squillace}, {Weinreb}, {Woody}, {Yadlapalli}, {Ahumada}, {Dong}, {Fremling}, {Huang}, {Karambelkar}, \& {Miller}}]{Ravi+23}
{Ravi}, V., {Catha}, M., {Chen}, G., {et~al.} 2023, \bibinfo{title}{{Deep Synoptic Array Science: Discovery of the Host Galaxy of FRB 20220912A},} \apjl, 949, L3, \dodoi{10.3847/2041-8213/acc4b6}

\bibitem[{E. {Scannapieco} \& L. {Bildsten}(2005){Scannapieco} \& {Bildsten}}]{Scannapieco+05}
{Scannapieco}, E., \& {Bildsten}, L. 2005, \bibinfo{title}{{The Type Ia Supernova Rate},} \apjl, 629, L85, \dodoi{10.1086/452632}

\bibitem[{S. {Schulze} {et~al.}(2021){Schulze}, {Yaron}, {Sollerman}, {Leloudas}, {Gal}, {Wright}, {Lunnan}, {Gal-Yam}, {Ofek}, {Perley}, {Filippenko}, {Kasliwal}, {Kulkarni}, {Neill}, {Nugent}, {Quimby}, {Sullivan}, {Strotjohann}, {Arcavi}, {Ben-Ami}, {Bianco}, {Bloom}, {De}, {Fraser}, {Fremling}, {Horesh}, {Johansson}, {Kelly}, {Kne{\v{z}}evi{\'c}}, {Kne{\v{z}}evi{\'c}}, {Maguire}, {Nyholm}, {Papadogiannakis}, {Petrushevska}, {Rubin}, {Yan}, {Yang}, {Adams}, {Bufano}, {Clubb}, {Foley}, {Green}, {Harmanen}, {Ho}, {Hook}, {Hosseinzadeh}, {Howell}, {Kong}, {Kotak}, {Matheson}, {McCully}, {Milisavljevic}, {Pan}, {Poznanski}, {Shivvers}, {van Velzen}, \& {Verbeek}}]{Schulze+21}
{Schulze}, S., {Yaron}, O., {Sollerman}, J., {et~al.} 2021, \bibinfo{title}{{The Palomar Transient Factory Core-collapse Supernova Host-galaxy Sample. I. Host-galaxy Distribution Functions and Environment Dependence of Core-collapse Supernovae},} \apjs, 255, 29, \dodoi{10.3847/1538-4365/abff5e}

\bibitem[{S. Seabold \& J. Perktold(2010)Seabold \& Perktold}]{statsmodels}
Seabold, S., \& Perktold, J. 2010, in 9th Python in Science Conference

\bibitem[{V. {Shah} {et~al.}(2025){Shah}, {Shin}, {Leung}, {Fong}, {Eftekhari}, {Amiri}, {Andersen}, {Andrew}, {Bhardwaj}, {Brar}, {Cassanelli}, {Chatterjee}, {Curtin}, {Dobbs}, {Dong}, {Dong}, {Fonseca}, {Gaensler}, {Halpern}, {Hessels}, {Ibik}, {Jain}, {Joseph}, {Kaczmarek}, {Kahinga}, {Kaspi}, {Kharel}, {Landecker}, {Lanman}, {Lazda}, {Main}, {Mas-Ribas}, {Masui}, {Mckinven}, {Mena-Parra}, {Meyers}, {Michilli}, {Nimmo}, {Pandhi}, {Patil}, {Pearlman}, {Pleunis}, {Prochaska}, {Rafiei-Ravandi}, {Sammons}, {Sand}, {Scholz}, {Smith}, \& {Stairs}}]{Shah+25}
{Shah}, V., {Shin}, K., {Leung}, C., {et~al.} 2025, \bibinfo{title}{{A Repeating Fast Radio Burst Source in the Outskirts of a Quiescent Galaxy},} \apjl, 979, L21, \dodoi{10.3847/2041-8213/ad9ddc}

\bibitem[{K. {Sharma} {et~al.}(2024){Sharma}, {Ravi}, {Connor}, {Law}, {Ocker}, {Sherman}, {Kosogorov}, {Faber}, {Hallinan}, {Harnach}, {Hellbourg}, {Hobbs}, {Hodge}, {Hodges}, {Lamb}, {Rasmussen}, {Somalwar}, {Weinreb}, {Woody}, {Leja}, {Anand}, {Das}, {Qin}, {Rose}, {Dong}, {Miller}, \& {Yao}}]{Sharma+24}
{Sharma}, K., {Ravi}, V., {Connor}, L., {et~al.} 2024, \bibinfo{title}{{Preferential occurrence of fast radio bursts in massive star-forming galaxies},} \nat, 635, 61, \dodoi{10.1038/s41586-024-08074-9}

\bibitem[{K. {Shin} {et~al.}(2023){Shin}, {Masui}, {Bhardwaj}, {Cassanelli}, {Chawla}, {Dobbs}, {Dong}, {Fonseca}, {Gaensler}, {Herrera-Mart{\'\i}n}, {Kaczmarek}, {Kaspi}, {Leung}, {Merryfield}, {Michilli}, {M{\"u}nchmeyer}, {Pearlman}, {Rafiei-Ravandi}, {Smith}, {Stairs}, \& {Tendulkar}}]{Shin+23}
{Shin}, K., {Masui}, K.~W., {Bhardwaj}, M., {et~al.} 2023, \bibinfo{title}{{Inferring the Energy and Distance Distributions of Fast Radio Bursts Using the First CHIME/FRB Catalog},} \apj, 944, 105, \dodoi{10.3847/1538-4357/acaf06}

\bibitem[{R.~E. {Skelton} {et~al.}(2014){Skelton}, {Whitaker}, {Momcheva}, {Brammer}, {van Dokkum}, {Labb{\'e}}, {Franx}, {van der Wel}, {Bezanson}, {Da Cunha}, {Fumagalli}, {F{\"o}rster Schreiber}, {Kriek}, {Leja}, {Lundgren}, {Magee}, {Marchesini}, {Maseda}, {Nelson}, {Oesch}, {Pacifici}, {Patel}, {Price}, {Rix}, {Tal}, {Wake}, \& {Wuyts}}]{Skelton+14}
{Skelton}, R.~E., {Whitaker}, K.~E., {Momcheva}, I.~G., {et~al.} 2014, \bibinfo{title}{{3D-HST WFC3-selected Photometric Catalogs in the Five CANDELS/3D-HST Fields: Photometry, Photometric Redshifts, and Stellar Masses},} \apjs, 214, 24, \dodoi{10.1088/0067-0049/214/2/24}

\bibitem[{S. {Tacchella} {et~al.}(2022){Tacchella}, {Conroy}, {Faber}, {Johnson}, {Leja}, {Barro}, {Cunningham}, {Deason}, {Guhathakurta}, {Guo}, {Hernquist}, {Koo}, {McKinnon}, {Rockosi}, {Speagle}, {van Dokkum}, \& {Yesuf}}]{Tacchella+22}
{Tacchella}, S., {Conroy}, C., {Faber}, S.~M., {et~al.} 2022, \bibinfo{title}{{Fast, Slow, Early, Late: Quenching Massive Galaxies at z {\ensuremath{\sim}} 0.8},} \apj, 926, 134, \dodoi{10.3847/1538-4357/ac449b}

\bibitem[{K. {Taggart} \& D.~A. {Perley}(2021){Taggart} \& {Perley}}]{Taggart&Perley21}
{Taggart}, K., \& {Perley}, D.~A. 2021, \bibinfo{title}{{Core-collapse, superluminous, and gamma-ray burst supernova host galaxy populations at low redshift: the importance of dwarf and starbursting galaxies},} \mnras, 503, 3931, \dodoi{10.1093/mnras/stab174}

\bibitem[{S.~P. {Tendulkar} {et~al.}(2017){Tendulkar}, {Bassa}, {Cordes}, {Bower}, {Law}, {Chatterjee}, {Adams}, {Bogdanov}, {Burke-Spolaor}, {Butler}, {Demorest}, {Hessels}, {Kaspi}, {Lazio}, {Maddox}, {Marcote}, {McLaughlin}, {Paragi}, {Ransom}, {Scholz}, {Seymour}, {Spitler}, {van Langevelde}, \& {Wharton}}]{Tendulkar+17}
{Tendulkar}, S.~P., {Bassa}, C.~G., {Cordes}, J.~M., {et~al.} 2017, \bibinfo{title}{{The Host Galaxy and Redshift of the Repeating Fast Radio Burst FRB 121102},} \apjl, 834, L7, \dodoi{10.3847/2041-8213/834/2/L7}

\bibitem[{P. Virtanen {et~al.}(2020)Virtanen, Gommers, Oliphant, Haberland, Reddy, Cournapeau, Burovski, Peterson, Weckesser, Bright, {van der Walt}, Brett, Wilson, Millman, Mayorov, Nelson, Jones, Kern, Larson, Carey, Polat, Feng, Moore, {VanderPlas}, Laxalde, Perktold, Cimrman, Henriksen, Quintero, Harris, Archibald, Ribeiro, Pedregosa, {van Mulbregt}, \& {SciPy 1.0 Contributors}}]{scipy}
Virtanen, P., Gommers, R., Oliphant, T.~E., {et~al.} 2020, \bibinfo{title}{{{SciPy} 1.0: Fundamental Algorithms for Scientific Computing in Python},} Nature Methods, 17, 261, \dodoi{10.1038/s41592-019-0686-2}

\bibitem[{Z. {Wadiasingh} \& A. {Timokhin}(2019){Wadiasingh} \& {Timokhin}}]{Wadiasingh&Timokhin19}
{Wadiasingh}, Z., \& {Timokhin}, A. 2019, \bibinfo{title}{{Repeating Fast Radio Bursts from Magnetars with Low Magnetospheric Twist},} \apj, 879, 4, \dodoi{10.3847/1538-4357/ab2240}

\bibitem[{Z. {Wang} {et~al.}(2025){Wang}, {Bannister}, {Gupta}, {Deng}, {Pilawa}, {Tuthill}, {Bunton}, {Flynn}, {Glowacki}, {Jaini}, {Lee}, {Lenc}, {Lucero}, {Paek}, {Radhakrishnan}, {Thyagarajan}, {Uttarkar}, {Wang}, {Bhat}, {James}, {Moss}, {Murphy}, {Reynolds}, {Shannon}, {Spitler}, {Tzioumis}, {Caleb}, {Deller}, {Gordon}, {Marnoch}, {Ryder}, {Simha}, {Anderson}, {Ball}, {Brodrick}, {Cooray}, {Gupta}, {Hayman}, {Ng}, {Pearce}, {Phillips}, {Voronkov}, \& {Westmeier}}]{Wang+25}
{Wang}, Z., {Bannister}, K.~W., {Gupta}, V., {et~al.} 2025, \bibinfo{title}{{The CRAFT coherent (CRACO) upgrade I: System description and results of the 110-ms radio transient pilot survey},} \pasa, 42, e005, \dodoi{10.1017/pasa.2024.107}

\bibitem[{E. {Waxman}(2017){Waxman}}]{Waxman17}
{Waxman}, E. 2017, \bibinfo{title}{{On the Origin of Fast Radio Bursts (FRBs)},} \apj, 842, 34, \dodoi{10.3847/1538-4357/aa713e}

\bibitem[{ {W}es {M}c{K}inney(2010){W}es {M}c{K}inney}]{pandas}
{W}es {M}c{K}inney. 2010, in {P}roceedings of the 9th {P}ython in {S}cience {C}onference, ed. {S}t\'efan van~der {W}alt \& {J}arrod {M}illman, 56 -- 61, \dodoi{10.25080/Majora-92bf1922-00a}

\bibitem[{M.~N. {Woodland} {et~al.}(2024){Woodland}, {Mannings}, {Prochaska}, {Ryder}, {Marnoch}, {Jorgenson}, {Simha}, {Tejos}, {Gordon}, {Fong}, {Kilpatrick}, {Deller}, \& {Glowacki}}]{Woodland+24}
{Woodland}, M.~N., {Mannings}, A.~G., {Prochaska}, J.~X., {et~al.} 2024, \bibinfo{title}{{The Environments of Fast Radio Bursts Viewed Using Adaptive Optics},} \apj, 973, 64, \dodoi{10.3847/1538-4357/ad643c}

\end{thebibliography}
\bibliographystyle{aasjournalv7}



\end{document}